\documentclass{amsart}
\usepackage{amsmath}
  \usepackage{paralist}
  \usepackage{graphics} 
  \usepackage{epsfig} 
\usepackage{graphicx}  \usepackage{epstopdf}
 \usepackage[colorlinks=true]{hyperref}
\hypersetup{urlcolor=blue, citecolor=red}

\usepackage{multirow}
\usepackage{pgffor}
\usepackage{longtable}
\usepackage{booktabs}

\def\currentvolume{X}
 \def\currentissue{X}
  \def\currentyear{200X}
   \def\currentmonth{XX}
    \def\ppages{X--XX}

\newtheorem{theorem}{Theorem}[section]

\theoremstyle{definition}

\newtheorem{method}{Method}

\usepackage{hyperref}
\usepackage{url}
\usepackage{booktabs}
\usepackage{bbold}
\usepackage{amsmath}
\DeclareMathOperator*\img{img}

\title[Multiple testing with persistent homology] 
      {Multiple testing with persistent homology}

\author[M. Vejdemo-Johansson and S. Mukherjee]{}


 \email{mvj@math.csi.cuny.edu}
 \email{sayan.mukherjee@mis.mpg.de}

\thanks{    MVJ would like to acknowledge the support of NVIDIA Corporation with the donation of the Titan X Pascal GPU used for this research.
    \newline
    SM would like to acknowledge partial funding from HFSP RGP005, NSF DMS 17-13012, NSF BCS 1552848, NSF DBI 1661386, NSF IIS 15-46331, NSF DMS 16-13261, as well as high-performance computing partially supported by grant 2016-IDG-1013 from the North Carolina Biotechnology Center
    \newline
    The simulated data used in this paper are available from Figshare, DOI 10.6084/m9.figshare.10262507
}

\thanks{$^*$ Corresponding author: M Vejdemo-Johansson}

\begin{document}
\maketitle

\centerline{\scshape Mikael Vejdemo-Johansson$^*$}
\medskip
{\footnotesize
  \centerline{CUNY College of Staten Island}
  \centerline{CUNY Graduate Center}
  \centerline{2800 Victory Boulevard, 1S-215}
  \centerline{Staten Island, NY 10314, USA}
} 

\medskip

\centerline{\scshape Sayan Mukherjee}
\medskip
{\footnotesize
\centerline{Center for Scalable Data Analytics and Artificial Intelligence},
\centerline{Universität Leipzig,
Humboldtstraße 25, Leipzig, Germany 04105} 
\centerline{Max Planck Institute for Mathematics
in the Sciences, Inselstraße 22 04103 Leipzig Germany} 
\centerline{Departments of Statistical Science, Mathematics, Computer Science, and Biostatistics \& Bioinformatics}
\centerline{Duke University}
\centerline{Durham, NC 27708, USA}
}

\bigskip

\begin{abstract}
In this paper we propose a computationally efficient multiple hypothesis testing procedure for persistent homology. The computational efficiency of our procedure is based on the observation that one can empirically simulate a null distribution that is universal across many hypothesis testing applications involving persistence homology. Our observation suggests that one can simulate the null distribution 
efficiently based on a small number of summaries of the collected data and use this null in the same way that p-value tables were used in classical statistics. To illustrate the efficiency and utility of the null distribution we provide procedures for rejecting acyclicity with both control of the Family-Wise Error Rate (FWER) and the False Discovery Rate (FDR). We will argue that the empirical null we propose
is very general conditional on a few summaries of the data based on simulations and  limit theorems for persistent homology for point processes.
\end{abstract}

\section{Introduction}
\label{sec:introduction}

Hypothesis testing based on topological summaries of data has been an area of Topological Data Analysis  (TDA)  that has seen growth recently. A challenge in applying the classical Neyman-Pearson hypothesis testing paradigm to topological summaries of data such as persistence homology is that the distribution of the test statistic under the null hypothesis is unknown. The issue of the lack of distribution theory for null hypotheses in TDA has been side-stepped
by the use of sampling and permutation based approaches to compute empirical null distributions. The computational cost of simulating these empirical null distributions from data has not been of concern in TDA as the number of tests have been small. For example, in the two sample testing case \cite{robinson_hypothesis_2017} there is one test and in the  analysis of variance (ANOVA) settings \cite{cericola_extending_2016} one typically considers a small number of conditions.

In this paper we consider multiple hypothesis testing (MHT) using topological summaries of data. One way MHT can arise is when the number of groups tested for in the ANOVA gets large. In addition, there are an increasing number of applications in which one tests
subsets of features in a persistence diagram with the goal of finding subsets of the data that significantly differ between the two groups. As in most feature selection problems MHT control is required in this setting. The computational cost simulating of data-driven empirical null distributions across thousands of tests can become prohibitive.
In many MHT conditions the null distribution for each test is known, for example each individual test may be a t-test and the null distribution is the student's t-distribution. The existence of an analytic formula of the distribution for a particular null model allows for efficient MHT and is an important notion of universality.

A recent idea that has begun to be explored in TDA is the question
of the universality of the distribution of certain topological summaries. For example, in \cite{bobskra} extensive simulations were
used to provide a great deal of evidence that the ratio of birth times to death times for noisy features in persistence diagrams is universal, independent of the model generating the point-cloud. In this paper we propose an efficient procedure to simulate an empirical null distribution that is universal. The advantage of this empirical null is one can compute this null model before observing data and use it across various models that may have generated the data. We will provide evidence for the universality of this empirical null based on simulations and central limit theorem based arguments. The empirical
null distribution we simulate is based on topological summaries of 
a homogeneous Poisson point process on $\mathbb{R}^d$ with unit intensity.  We will argue for the general case use of this null distribution based on simulations as well as central limit theorems proven in~\cite{hiraoka_limit_2018}. The idea is that we can apply this distribution irrespective of model generating the data. We also provide both Family-Wise Error Rate (FWER) and False Discovery Rate (FDR) control procedures using the null distribution we propose.

\subsection{Structure of this paper}
\label{sec:structure-this-paper}

Section~\ref{sec:tda-background} introduces basic concepts from Topological Data Analysis (TDA) and  subsection \ref{sec:statistics-persistence} provides an overview of existing work on hypothesis testing in TDA. Section~\ref{sec:noise-model-one} introduces our proposed null model, the procedure to generate the
empirical null distribution, as well as the hypothesis testing procedure. Section~\ref{sec:fwer-fdr} introduces the multiple hypothesis testing procedures
we propose. Section~\ref{sec:comparing-statistics} provides a justification based on distribution theory for the empirical null distribution we propose. This argument is based on a central limit theorem for persistent homology provided in~\cite{hiraoka_limit_2018}  Section~\ref{sec:experiment-setup} describes the simulations that form the basis for our experimental validation and statistical power measurement experiments, and Section~\ref{sec:experiments-results} contain the experiments using the data generated in the simulations from the preceding Section.

\section{TDA background: persistent homology and hypothesis testing}
\label{sec:tda-background}

In this paper, we concern ourselves with the statistical behavior of \emph{persistent homology}.

An \emph{abstract simplicial complex} is a collection $\Sigma$ of subsets, called simplices, of a set $V$ of \emph{vertices} such that all subsets $\tau\subset\sigma\in\Sigma$ of a simplex are also simplices in $\Sigma$.
A simplex with $d+1$ vertices is said to be $d$-dimensional.
To a simplicial complex is associated the \emph{chain complex}: a graded vector space $C_*\Sigma$ with basis elements corresponding to the simplices in $\Sigma$.
The grading is provided by the dimension of the corresponding simplex.
The chain complex is equipped with a linear \emph{boundary map}:
\[
\partial([v_0,\dots,v_d]) = \sum(-1)^i[v_0,\dots,v_{i-1},v_{i+1},\dots,v_d]
\]
The \emph{homology} $H_*\Sigma$ is defined as $\ker\partial/\img\partial$.

For a comprehensive introduction to homology we refer to Hatcher \cite{hatcher_algebraic_2002}.

A simplicial complex is \emph{filtered} if it decomposes into a sequence of inclusions $\Sigma_s\hookrightarrow\Sigma_t$ for $s<t$.
\emph{Persistent homology} provides a way of gluing together the individual homology vector spaces $H_*\Sigma_t$ into a globally consistent structure $PH_*\Sigma_*$.
In this structure, the bases for each $H_*\Sigma_t$ are chosen to be compatible: each basis element emerges at some $t_b$ and is preserved through all parameter values up to some $t_d$. 
The persistent homology is often represented as a \emph{persistence diagram}, a multiset in the plane with each basis element represented as its pair of birth and death times: $\{(t_b,t_d)\}$. We will be interested in testing statistical properties of birth-vs-death times, for example differences or ratios.

The class of persistence diagrams admits several interesting metrics. The most commonly used in TDA is the \emph{Bottleneck distance}. Each diagram 
is a multiset of points in the extended plane where in addition to the  finitely many off-diagonal points, each diagram includes copies of all points on
the diagonal. The diagonal is needed so that bijections can be defined between diagrams that have different number of off-diagonal points. For a 
pair $X, Y$ of persistence diagrams the \emph{bottleneck distance} is defined as
The \emph{bottleneck distance} is defined as
\[
d_{\mathcal B}(X,Y) = \inf_\phi \sup_x\|x-\phi(x)\|_\infty,
\]
where $\phi$ is the set of all possible bijections between  $X$ and  $Y$. 

\subsection{Hypothesis testing with persistent homology}
\label{sec:statistics-persistence}

The idea that topological summaries such as persistence diagrams form a probability space for which formal statistical analysis is well defined was developed in \cite{mileyko_2011}. 
Further developments on defining useful summary statistics within persistent homology and considering means, medians, and variances of persistence diagrams was pursued in several papers \cite{munch_probabilistic_2015,turner_means_2013,turner_sufficient_2013}. The main challenge in considering persistence diagrams as a probability space was pointed out in
\cite{turner_means_2013,turner_frechet_2014}---the space of persistence diagrams is positively curved which results in non-unique geodesics. As a result the mean of a set of diagrams need not be unique which complicates data analysis. To avoid this issue persistence landscapes were introduced in \cite{bubenik_statistical_2015}, persistence landscapes are functions so they can be considered as random functions in a Banach space, a construction that admits  central limit theorems, unique means and medians. Further examination of bootstrap properties of persistence based summaries as well as a notion of confidence intervals for points in a diagram was developed in \cite{chazal_bootstrap_2013,chazal_stochastic_2013}.
An alternative approach was considered in a series of papers where instead of considering a persistence diagram as a summary a probability density was used as a topological summary, an approach called distance to measure \cite{chazal_robust_2014,chazal_subsampling_2014,chazal_bootstrap_2013}

A theoretical basis for hypothesis testing using topological summaries including persistence diagrams was 
outlined in \cite{blumberg_robust_2014}. The authors advocated for the use of empirical distributions of persistence diagrams and topological summaries. The authors denote as $\Phi_k^n(X,\mu_X)$ a collection of $k$ topological summaries of $n$ points drawn from a metric space $X$ with measure $\mu_X$. They stated that the empirical measure $\Phi_k^n(X,\mu_X)$ is close to the empirical measure $\Phi_k^n(S)$ where $S$ is a large sample, $|S| \gg n$, where $k$ subsamples of size $n$ are drawn from $S$. The empirical measure $\Phi_k^n(S)$ can be computed and used for hypothesis testing, as well as computing significance intervals. Note, here again the idea of an empirical measure constructed in a data-driven way is central to the hypothesis testing procedure. Hypothesis testing procedures based on permutation tests and barcode distances for
two sample testing \cite{robinson_hypothesis_2017} and ANOVA ~\cite{cericola_extending_2016} can both be justified under the framework in 
\cite{blumberg_robust_2014}. Our approach for MHT also can be formulated using the same theoretical framework, except we argue that the null distributions conditioned on different base observations can be compared with each other for the tests we consider.

\section{Multiple hypothesis testing with topological summary statistics}
\label{sec:MHT}

In this section we outline procedures for for multiple hypothesis testing with topological summary statistics. We will use testing for acyclicity as an example application for our procedure.  We start with the single test one-sample setting to fix notation and provide a base procedure in subsection \ref{sec:noise-model-one}. We then show how we can standardize the null distribution across tests for topological summaries in subsection \ref{sec:comparing-statistics}, this standardization is typically required for multiple testing. We close this section by stating MHT correction procedures for controlling the familywise error rate or the false discovery rate for both one-sample and two-sample tests
in subsection \ref{sec:fwer-fdr}.

\subsection{Null model and a one-sample testing procedure}
\label{sec:noise-model-one}

We are interested in the following question given a large collection of topological statistics computed from an observed point cloud $\{x_1,...,x_n\}$ in $\mathbb{R}^d$ is there ``interesting topology''.
From a hypothesis testing framework we need first need a null hypothesis and an alternative hypothesis
\begin{equation} \label{nullalt}
H_0: \{x_1,...,x_n\} \stackrel{iid}{\sim} \mathcal{P}_0, \quad H_A: \{x_1,...,x_n\} \stackrel{iid}{\sim} \mathcal{P}_A,
\end{equation}
where under the null model $\mathcal{P}_0$ is the uniform distribution with the support of a convex body in $\mathbb{R}^d$ and the alternative model is non-uniform distribution supported on the same convex body in $\mathbb{R}^d$. The null model captures topological summaries of a uniform uniform distribution on a convex body, what we consider a signal is deviation from the uniform distribution. We are not interested in homological obstructions characterizing the null hypothesis.



\subsubsection{The null distribution and the testing procedure} 
\label{sec:null-distributions-one-sample-test}
We will state a procedure for generating the null distribution and our one sample test procedure. We assume the point cloud we obtain $X=\{x_1,...,x_n\}$ is drawn from a distribution with support on a convex body in $\mathbb{R}^d$. Given
a point cloud we compute a diagram $\mathcal{D}$ and a test statistic
from the diagram denoted as $\gamma(\mathcal{D})$.
Given a collection of birth-death points $\{b_i,d_i\}_{i=1}^N$
examples of test statistics we consider are
$\gamma = \max_{i=1,...N}(d_i-b_i)$,
$\gamma = \log\max_{i=1,...N}(d_i-b_i)$, $\gamma = \sum_{i=1,...N}(d_i-b_i)$; $\gamma = \max_{i=1,...N}(d_i/b_i)$,
$\gamma = \log\log\max_{i=1,...N}(d_i/b_i)$, $\gamma = \sum_{i=1,...N}(d_i/b_i)$.

The basic idea is that our null model for a point cloud is a uniform distribution supported on a convex body. We would like the convex body of the null model to match as best as possible the support of the distribution generating the data, which we also assume is convex. 
We propose two options for matching the convex body of the null model to support of the distribution on the observed data. The more computationally costly procedure is to compute the convex hull of our data $\mathcal{C} = \mbox{Conv}(X=\{x_1,...,x_n\})$ and use $\mathcal{C}$ as the support of of
the density in our null model $H_0$. The computationally cheaper procedure computes a bounding box from the data $\mathcal{B} = \mbox{Box}(X=\{x_1,...,x_n\})$, a bounding box provides intervals in each axis-parallel direction as the support. Note that both $\mathcal{B}$
and $\mathcal{C}$ are convex bodies. The requirement of a convex body arises from needing to standardize the null distributions across the multiple hypothesis, we discuss this in detail in \ref{sec:comparing-statistics}.


The convex hull procedure is computational much more expensive than the bounding box procedure due to having to both compute the convex hull as well as
sample uniformly from a convex hull. 

Another issue with the convex hull procedure is that without adjustments, the procedure is likely to be biased: just like in point process statistics, most point clouds will be unlikely to have points \emph{on} the boundary, while the convex hull procedure is guaranteed to have several points on the boundary (in fact, defining the boundary).

In \cite{moore-1984}, the work by \cite{ripley-rasson-1977,rasson-1979} is described and generalized, giving estimators for convex hulls in the plane with a straightforward formula and for higher-dimensional convex hulls with a Bayesian decision theoretic approach. The construction calls for a convex hull $H(x)$ of a sample $x$ drawn from a convex shape $D$ to be translated by the centroid $c(H(x))$ to $H(x)-c(H(x))$, and then dilated by a factor $\sqrt{m(D)/m(H(x))}$ for the Lebesgue measure $m$ of the shape $D$ and the convex hull. Since $m(D)$ is often unknown, an unbiased estimator of the dilation factor is $c=\sqrt{(n+1)/(n+1-\mathbb{E}[V_{n+1}]}$ where $V_{n+1}$ is the number of points on the boundary of the convex hull after drawing one additional point from $D$. \cite{rasson-1979} has a complicated expression under certain conditions for $\mathbb{E}[V_{n+1}]$, but \cite{moore-1984} recommends instead to estimate $\mathbb{E}[V_{n+1}]$ using the observed count of boundary points $v_n$. In summary, an unbiased estimator of the convex hull would be dilating the computed convex hull by a factor of $c=\sqrt{n/(n-v_n)}$.

For the bounding box procedure there is a simple uniformly minimum variance unbiased estimator of the bounding box first studied by \cite{lloyd-1952}
by computing for each coordinate $i$ the upper and lower bound of the box as
\[
{\hat a}_i = \frac{n\min_i - \max_i}{n-1}
\qquad
{\hat b}_i = \frac{n\max_i - \min_i}{n-1}
\]
where $\mbox{min}_i$ is the smallest value in coordinate $i$ and  $\mbox{max}_i$ is the largest value. 
Our recommendation is to use the unbiased convex hull procedure.

For the computation of the $p$-values of a simulation test, we follow the advice in \cite{phipson2010permutation} and estimate $\hat{p} = (\#\{\gamma_\pi<\gamma^*\}+1)/(\Pi+1)$.

\begin{method}[Test procedure]
\label{mth:one-sample-simulation}
-- Given the point cloud data $X=\{x_1,...,x_n\}$ as input.
\begin{enumerate}
    \item[(1)] Compute the diagram $\mathcal{D}(X)$ and statistic  $\gamma^* = \gamma(\mathcal{D}(X))$.
    \item[(2)] Compute the convex body $\mathcal{V}$ as either $\mathcal{C}$ or $\mathcal{B}$.
    \item[(3)] For $\pi = 1,...,\Pi$ 
    \begin{enumerate}
        \item[(a)] Sample a point cloud $X_\pi$ of $n$ points sampled independently from the uniform distribution over the convex hull,
        $x_{\pi,1},...,x_{\pi,n} \stackrel{iid}{\sim} \mbox{U}(\mathcal{V})$.
        \item[(b)] Compute the diagram $\mathcal{D}(X_\pi)$ and the statistic of interest $\gamma_\pi = \gamma(\mathcal{D}(X_\pi))$.
    \end{enumerate}
    \item[(4)] Compute the p-value
    \begin{enumerate}
        \item[(a)] Left-tailed: p-value =$\frac{\#\{\gamma_\pi < \gamma^*\}+1}{\Pi+1}$
        \item[(b)] Right-tailed: p-value =$\frac{\#\{\gamma_\pi > \gamma^*\}+1}{\Pi+1}$
        \item[(c)] Two-tailed: p-value = $\min\left(\frac{\#\{\gamma_\pi < \gamma^*\}+1}{\Pi+1}, \frac{\#\{\gamma_\pi > \gamma^*\}+1}{\Pi+1}\right)$
    \end{enumerate}
\end{enumerate}
\end{method}

\subsection{Comparing persistence statistics -- standardizing tests}
\label{sec:comparing-statistics}

A key assumption in most MHT procedures is each test statistic shares the same null distribution. This situation may not be the case for topological statistics, 
for example differences
in the scale of the point cloud may cause topological
invariants to be not be immediately comparable.
We are considering statistics
of the form
$$f(t_d,t_b) = g(t_d-t_b)$$
where $g$ is a monotone function. Two examples of statistics we consider are
$$ \gamma = (t_d-t_b)/\sqrt2, \quad \gamma = \log((t_d-t_b)/\sqrt2).$$

We propose a very simple standardization procedure across the tests, we also provide a theoretical for our normalization procedure.
For each test we center and scale the test statistics the test statistics $\{\gamma_1,...,\gamma_\Pi\}$ as 
$$\gamma_{\pi} \mapsto \frac{\gamma_\pi- \widehat{\mu}_\pi}{\widehat{\sigma}_\pi},
\quad \widehat{\mu}_\pi = \frac{1}{\Pi} \sum_{\pi=1}^\Pi \gamma_\pi, \quad \widehat{\sigma}_\pi = \frac{1}{\Pi-1} \sum_{\pi=1}^\Pi(\gamma_\pi-\widehat{\mu}_\pi)^2.$$ We will show via simulations that this standardization allows us to map topological statistics to very similar null distributions
across tests.

Another argument for the z-score type standardization we proposed above is based on a central limit theorem stated in \cite{hiraoka_limit_2018}.
Consider $L$ as a convex shape and $\ell L$ a scaling of the shape. $\Phi_L$ is a stationary point process with its support restricted to $\ell L$. Following \cite{hiraoka_limit_2018}, we write $\mathbb{K}(\Phi_{ L})$ for a filtered simplicial complex constructed from a draw from $\Phi_L$ and  $\beta^{r,s}_q(\mathbb{K}(\Phi_{ L}))$ for the persistent dimension $q$ Betti number from $r$ to $s$ ($0  \leq r \leq s < \infty$) generated from the filtration $\mathbb{K}(\Phi_{ L})$. In other words, $\beta^{r,s}_q(\mathbb{K}(\Phi_L)$ counts the number of $q$-dimensional persistent homology features from the point cloud generated by $\Phi$ restricted to $L$ that are born before $r$ and survive past $s$. The following theorem suggests a z-score normalization.

\begin{theorem}\label{thm:convex-hiraoka-clt}[\cite{hiraoka_limit_2018}, Theorem 5.2]
Assume that $\Phi$ is a stationary point process having all finite moments. Then, for any $0\leq r\leq s<\infty$ and $q\geq 0$, 
and any convex shape $L$
\[
\frac{\beta_q^{r,s}(\mathbb{K}(\Phi_{\ell L})) -  \mathbb{E}\left[\beta_q^{r,s}(\mathbb{K}(\Phi_{\ell L}))\right]}{n^{1/2}}
\stackrel{d}{\rightarrow} \mbox{N}(0,\sigma^2_{r,s}), \quad \mbox{ as } n \rightarrow \infty.
\]
\end{theorem}
The above theorem states that there is a limiting CLT and so centering and scaling  distributions of persistence Betti 
numbers should result in a universal distribution, here the normal distribution. Empirically we observe that for some point 
cloud data we do not have the sample size to assume normality so we simulate an empirical null. An interesting question is to
better understand these asymptotics.


\subsection{Controlling for multiple testing}
\label{sec:fwer-fdr}

In the classical Neyman-Pearson paradigm for hypothesis testing one either \emph{rejects the null hypothesis} or \emph{fails to reject the null hypothesis} 
based on thresholds on the test statistic based on an $\alpha$ level of spuriously rejecting the null when the null is true. Consider not one but $m=10\,000$ hypothesis tests with an $\alpha$ level
of $.05$ to reject the null hypothesis, one would expect that the null hypothesis will be incorrectly rejected in $500$ of the tests.
This raises the need to control the number of false positives as one may need to conduct experiments to validate these $500$ tests
In this section we will provide procedures to implement the two classical multiple hypothesis correction procedures: False Discovery Rate (FDR) corrections
\cite{Korthauerrev} and the Family Wise Error Rate (FWER) corrections \cite{Nicholsrev}. We will apply these corrections to testing 
for interesting acyclicity in data.

The quantities in the following table will be used to define both the FWER and FDR control procedures.
We are performing $m$ different hypothesis tests from which $m_0$ follow the null as the true distribution and $m_1$ have the alternative as true.
We reject $R$ and accept $W$ of the tests. The error quantities in the table are $V$ which quantifies the false discoveries (false positives, type I errors) and $T$ quantifies the missed discoveries (false negatives, type II errors).

\vspace{.1in}

\begin{center}
  \begin{tabular}{r|c|c|l}
  \hline
& Accept null & Reject null & Total \\ \hline
Null true & $U$ & $V$ & $m_0$ \\
Null false & $T$ & $S$ & $m_1$ \\ \hline
& $W$ & $R$ & $m$ \\
\hline
\end{tabular}  
\end{center}
\vspace{.1in}



In standard (single) hypothesis testing the $\alpha$-level provides control of false positives 
$\alpha = \mathbb{P}(V>0)$. 
There are two main extensions
to control false positives: \\
 (a) The  \textbf{family-wise error rate} is defined as $\mathrm{FWER}=\mathbb{P}(V>0)$, the probability making even a single false discovery over all of the multiple hypotheses. The FWER can be hard to control in many situations, it may be impossible to have zero falase positives.\\
(b) The \textbf{false discovery rate} is defined as the percentage of false discoveries amongst the tests that reject the null: $\mathrm{FDR}=\mathbb{E}[V/R \mid R >0]$. If there are no rejections then the ratio is set to $0$. Controlling for the FDR relaxes the strict control in the FWER by allowing for a percentage of false positives.

\subsubsection{Family-wise error rate control}
\label{sec:fwer}

Classical FWER control methods specify a stricter cutoff to simultaneously control the probability of a false positive across all the tests. The simplest procedure is the \emph{Bonferroni correction}~\cite{bonferroni1950sulle}
which is based on the union bound, the probability of a union of events is upper bounded by the sum of their probabilities.
If the events are independent the union bound is tight, alas in almost all testing problems the events are not independent and the Bonferroni correction is overly conservative. The Bonferroni correction adjusts the $\alpha$-level to reject at
$\alpha_{\mathrm{FWER}} = \alpha/m$,
with $\alpha$ as the desired bound of the probability of a false positive across all the tests. There has been a great deal of work to control FWERs that are far less conservative.  Holm's step-down and Hochberg's step-up procedures~\cite{hochberg_1988,holm1979simple} are commonly used alternatives to the Bonferroni correction -- these procedures 
also use a more stringent $\alpha$-level threshold across all tests but allow for dependencies between tests.


If a permutation or simulation setting is required to compute the null distribution for each test the number $m$ of tests can drive up the number of permutations or simulations required for an acceptable test level dramatically. If computations are expensive -- such as with persistence diagrams -- the complexity quickly becomes prohibitive.

In a previous conference paper~\cite{vejdemo-johansson_certified_2018} a method for controlling the FWER when testing for acyclicity in a dataset
was proposed. The proposed procedure in \cite{vejdemo-johansson_certified_2018} assumed a null hypothesis of the point cloud being generated from a uniform distribution. The procedures in this paper are further developments and refinements of the core idea in \cite{vejdemo-johansson_certified_2018}.

For the remainder of this paper we will consider the particular example of testing for acyclicity in a dataset to allow us to focus on a particular topological test. Adapting the procedures to other topological tests is straightforward.


The approach is rooted in the observation that, having computed standardized test statistics $(t_1,\dots,t_m)$ from each test separately,
\begin{align*}
\alpha_{\mathrm{FWER}} &= \mathbb{P}(V>0) \\
 &= \mathbb{P}(\{t_1 > c\} \cup \dots \cup \{t_m > c\} \mid H_0) \\
 &= \mathbb{P}(\max_i t_i > c \mid H_0)
\end{align*}
We do not need distributional assumptions as long as the null distributions are comparable across all tests. See subsection \ref{sec:comparing-statistics} for a simple standardization procedure.

We now state our proposed FWER correction procedure for testing for acyclicity:
\begin{method}[FWER correction procedure]\label{mth:fwer}
A collection of point clouds $(X_1,\dots,X_K)$;
a procedure to compute diagrams $\mathcal{D}(X)$;
a scalar test summary
statistic $\gamma$
computed from a diagram; a null model $\mathcal M$ and procedure to generate random point clouds;
the following procedure is a FWER correction to testing for acyclicty:
\begin{enumerate}
\item Draw $(M_1^1,\dots,M^N_K) \stackrel{iid}{\sim} \mathcal M$.
\item Compute diagrams $(\mathcal{D}_j(X_j))_{j=1}^K$ and $(\mathcal{D}^i_j)_{i=1,...,M,j=1,...,K}$.
\item Compute the statistics $\tilde y_j = \gamma(\mathcal{D}_j)$, $j=1,...,K$ and $\tilde y^i_j = \gamma(\mathcal{D}^i_j)$,
$i=1,...,M,j=1,..,K$.
\item For each $i=1,...,K$
compute $\hat{\mu}_i = \frac{1}{N} \sum_{j=1}^N \tilde y^i_j$ and $\hat{\sigma}_i = \frac{1}{N-1} \sum_{j=1}^N (\tilde y^i_j-\hat{\mu}_i)^2$
\item Standardize
$y_i := \frac{\tilde y_i - \hat{\mu}_i}{\hat{\sigma}_i}$ and
$y_i^j := \frac{\tilde y_i^j - \hat{\mu}_i}{\hat{\sigma}_i}$.. 
\item Compute $z_i=\max_j y_j^i$ for $i=1,...,K$ 
\end{enumerate}
We may then reject the null hypothesis at a level of $p=(\#\{z_i>y_i\}+1)/(N+1)$.
\end{method}

Picturing the observed $\tilde y_j^1$ as a column vector, this procedure completes the observed data to a full matrix by augmenting with a row of simulated $\tilde y_j^i$ for each observation. The simulated values in each row is used to standardize the entire row. Finally, each column is taken as an aggregate simulation and compared to the other columns in the final calculation.

\subsubsection{False discovery rates}
\label{sec:fdr}

For false discovery rate control, we seek to control  $q_{\mathrm{FDR}}=\mathbb{E}[V/R]$: the proportion of 
false discoveries among all the rejected null hypotheses. By convention, if $R=0$ then $q_{\mathrm{FDR}}=0$.

We now state our proposed FDR correction procedure for testing for acyclicity:
Based on this we propose the following approach
\begin{method}[FDR correction procedure]\label{mth:fdr}
A collection of point clouds $(X_1,\dots,X_K)$;
a procedure to compute diagrams $\mathcal{D}(X)$;
a scalar test summary
statistic $\gamma$
computed from a diagram; a null model $\mathcal M$ and procedure to generate random point clouds;
the following procedure is a FDR correction to testing for acyclicty:
\begin{enumerate}
\item Draw $(M_1^1,\dots,M^N_K) \stackrel{iid}{\sim} \mathcal M$.
\item Compute diagrams $(\mathcal{D}_j(X_j))_{j=1}^K$ and $(\mathcal{D}^i_j)_{i=1,...,M,j=1,...,K}$.
\item Compute the statistics $\tilde y_j = \gamma(\mathcal{D}_j)$, $j=1,...,K$ and $\tilde y^i_j = \gamma(\mathcal{D}^i_j)$,
$i=1,...,M,j=1,..,K$.
\item For each $i=1,...,K$
compute $\hat{\mu}_i = \frac{1}{N} \sum_{j=1}^N \tilde y^i_j$ and $\hat{\sigma}_i = \frac{1}{N-1} \sum_{j=1}^N (\tilde y^i_j-\hat{\mu}_i)^2$
\item Standardize
$y_i := \frac{\tilde y_i - \hat{\mu}_i}{\hat{\sigma}_i}$ and
$y_i^j := \frac{\tilde y_i^j - \hat{\mu}_i}{\hat{\sigma}_i}$.. 
\item Rank $(y_{(1)},...,y_{(K)})$ to form a sorted sequence.
\item For a fixed cutoff $c_i$, calculate
\[
\%V(c_i) = \frac{\#\{y_i^j \geq c_i : j>1\}}{K(N-1)}
\qquad
\%R(c_i) = \frac{\#\{x_i\geq c_i\}}{K}
\qquad
\hat q_{\mathrm{FDR}}(c_i) = \frac{\%V(c_i)}{\%R(c_i)}
\]
\item Pick the smallest $c_i$ such that $\hat q_{\mathrm{FDR}(c_i)} \leq \alpha$ for the chosen level $\alpha$.
\item Reject all null hypotheses where the test statistic is $\geq c_i$
\end{enumerate}
\end{method}

If all $\hat q_{\mathrm{FDR}}(c_i)\geq\alpha$, then this means that $\min_i\hat q_{\mathrm{FDR}}(c_i)$ is as good a false discovery rate as is attainable.

\subsection{Two-sample FDR controlled testing}
\label{sec:two-sample-fdr}

In \cite{robinson_hypothesis_2017}, Robinson and Turner propose a hypothesis test for persistent homology.
In their setup, two groups of persistence diagrams $(D_{1,1},\dots,D_{1,n})$ and $(D_{2,1},\dots,D_{2,m})$ are sampled, and using a loss function built from in-group $p$-Wasserstein distance
 \begin{eqnarray*}
F_{p,q}(\{D_{1}\}, \{D_{2}\}) &=& 
\frac{1}{2n(n-1)}\sum\sum d_p(D_{1,i},D_{1,j})^q + \\
& & \frac{1}{2m(m-1)}\sum\sum d_p(D_{2,i},D_{2,j})^q
\end{eqnarray*}
they propose a permutation test: the membership in group 1 or 2 is repeatedly permuted and the loss function computed for each permutation.
The rank of the loss as compared to the loss on the point cloud data produces a $p$-value estimate for the test.

In a follow-up paper, Cericola et al~\cite{cericola_extending_2016} propose an extension to Robinson and Turner's two sample test to test for an ANOVA style hypothesis of several groups 
$$D_{1,1},\dots,D_{1,n_1},\dots,D_{m,1},\dots,D_{h,n_m}$$ 
being all equal.
This paper suggests that after rejecting this type of null hypothesis, we can use Robinson and Turner style testing pairwise on the diagram groups to locate the discrepancies.

Neither of these two papers mention how to correct for the intrinsic multiple testing. We assume that a total of $h$ two sample tests are being performed.
Hence we are given $2h$ groups of persistence diagrams 
\begin{align*}
(D^1_{1,1},\dots,D^1_{1,n_1},&D^1_{2,1}\dots D^1_{2,m_1})\\
&\vdots \\
(D^h_{1,1},\dots,D^h_{1,n_h},&D^h_{2,1}\dots D^h_{2,m_h}). \\
\end{align*}

Following the basic philosophy used in Section~\ref{sec:fdr}, we propose the following procedure. \\
\begin{method}[False discovery error rate control for multiple 2-sample testing]\label{mth:multi-fdr} -- 
Given $h$ two sample tests as input.
\begin{enumerate}
    \item[(1)] Compute the $h$ distances $d^*_k = F_{p,q}(\{D^k_{1,i}\}_{i=1}^{n_k}, \{D^k_{2,j}\}_{j=1}^{m_k})$ $k=1,...,h$.
    \item[(2)] For $\pi = 1,...,\Pi$ 
    \begin{enumerate}
        \item[(a)] Permute the labels of all $h$ two sample tests
        \item[(b)] Compute the $h$ distances on the permuted data $d_{\pi,k} = F_{p,q}(\{D^{\pi,k}_{1,i}\}_{i=1}^{n_k}, \{D^{\pi,k}_{2,j}\}_{j=1}^{m_k})$ $k=1,...,h$.
            \end{enumerate}
    \item[(4)] Pick a FDR threshold $c$ and compute 
    \[
\%V(c) = \frac{\#\{d_{\pi,k} \geq c\}}{h \Pi}
\qquad
\%R(c) = \frac{\#\{d^*_k \geq c\}}{K}.
\]
 \item[(5)] The point clouds with $d^*_k \geq c$ have a FDR of  
$$q_{\mathrm{FDR}}(c) = \frac{\%V(c)}{\%R(c)}.$$
\end{enumerate}
\end{method}

Just like with Method~\ref{mth:fdr}, the cutoff $c$ is chosen as the smallest cutoff that still realizes the FDR required.

\subsection{Combining different invariants}
\label{sec:combining-invariants}

Since the test statistics are standardized separately for each point cloud, and since -- as we examine in Section \ref{sec:exchangeability} -- the resulting distributions of the standardized test statistics follow approximately the same distribution, even across different homological dimensions and different test statistic choices, we can adapt the FWER and FDR correction procedures above to not only address hypotheses that span different point clouds, but also to address hypotheses that span different test statistics.

This suggests the following modifications of Methods \ref{mth:fwer} and \ref{mth:fdr}:

\begin{method}[FWER correction with varying topological invariants]
\label{mth:fwer-varying}
A collection of point clouds $(X_1,\dots,X_K)$;
a procedure to compute diagrams $\mathcal{D}(X)$;
a scalar test summary
a collection of test statistics $(\gamma_1,\dots,\gamma_K)$
computed from a diagram; a null model $\mathcal M$ and procedure to generate random point clouds;
the following procedure is a FWER correction to testing for acyclicty:
\begin{enumerate}
\item Draw $(M_1^1,\dots,M^N_K) \stackrel{iid}{\sim} \mathcal M$.
\item Compute diagrams $(\mathcal{D}_j(X_j))_{j=1}^K$ and $(\mathcal{D}^i_j)_{i=1,...,M,j=1,...,K}$.
\item Compute the statistics $\tilde y_j = \gamma_j(\mathcal{D}_j)$, $j=1,...,K$ and $\tilde y^i_j = \gamma_j(\mathcal{D}^i_j)$,
$i=1,...,M,j=1,..,K$.
\item For each $i=1,...,K$
compute $\hat{\mu}_i = \frac{1}{N} \sum_{j=1}^N \tilde y^i_j$ and $\hat{\sigma}_i = \frac{1}{N-1} \sum_{j=1}^N (\tilde y^i_j-\hat{\mu}_i)^2$
\item Standardize
$y_i := \frac{\tilde y_i - \hat{\mu}_i}{\hat{\sigma}_i}$ and
$y_i^j := \frac{\tilde y_i^j - \hat{\mu}_i}{\hat{\sigma}_i}$.. 
\item Compute $z_i=\max_j y_j^i$ for $i=1,...,K$ 
\end{enumerate}
We may then reject the null hypothesis at a level of $p=(\#\{z_i>y_i\}+1)/(N+1)$.
\end{method}

\begin{method}[FDR correction with varying topological invariants]
\label{mth:fdr-varying}
A collection of point clouds $(X_1,\dots,X_K)$;
a procedure to compute diagrams $\mathcal{D}(X)$;
a scalar test summary
a collection of test statistics $(\gamma_1,\dots,\gamma_K)$
computed from a diagram; a null model $\mathcal M$ and procedure to generate random point clouds;
the following procedure is a FDR correction to testing for acyclicty:
\begin{enumerate}
\item Draw $(M_1^1,\dots,M^N_K) \stackrel{iid}{\sim} \mathcal M$.
\item Compute diagrams $(\mathcal{D}_j(X_j))_{j=1}^K$ and $(\mathcal{D}^i_j)_{i=1,...,M,j=1,...,K}$.
\item Compute the statistics $\tilde y_j = \gamma_j(\mathcal{D}_j)$, $j=1,...,K$ and $\tilde y^i_j = \gamma_j(\mathcal{D}^i_j)$,
$i=1,...,M,j=1,..,K$.
\item For each $i=1,...,K$
compute $\hat{\mu}_i = \frac{1}{N} \sum_{j=1}^N \tilde y^i_j$ and $\hat{\sigma}_i = \frac{1}{N-1} \sum_{j=1}^N (\tilde y^i_j-\hat{\mu}_i)^2$
\item Standardize
$y_i := \frac{\tilde y_i - \hat{\mu}_i}{\hat{\sigma}_i}$ and
$y_i^j := \frac{\tilde y_i^j - \hat{\mu}_i}{\hat{\sigma}_i}$.. 
\item Rank $(y_{(1)},...,y_{(K)})$ to form a sorted sequence.
\item For a fixed cutoff $c_i$, calculate
\[
\%V(c_i) = \frac{\#\{y_i^j \geq c_i : j>1\}}{K(N-1)}
\qquad
\%R(c_i) = \frac{\#\{x_i\geq c_i\}}{K}
\qquad
\hat q_{\mathrm{FDR}}(c_i) = \frac{\%V(c_i)}{\%R(c_i)}
\]
\item Pick the smallest $c_i$ such that $\hat q_{\mathrm{FDR}(c_i)} \leq \alpha$ for the chosen level $\alpha$.
\item Reject all null hypotheses where the test statistic is $\geq c_i$
\end{enumerate}
\end{method}

By repeating the same point cloud in several positions among the $X_1,\dots,X_K$, we can use these expanded methods to, for instance, construct acyclicity tests that span across all homological dimensions of interest: set $X_1=\dots=X_K=X$, and pick for your invariants some specific persistence diagram invariant applied in turn to the Betti-$d$ diagrams as $d$ ranges from $1$ to $K$.

\section{Simulations}
\label{sec:experiment-setup}

In this section we validate our procedures on simulations, testing for
single-sample acyclicity testing with FWER and FDR corrections. We will evaluate the performance of our tests using power analysis and provide evidence for the universality of our procedure by generating the alternative
hypothesis on a family of acyclic data.

In this section we first state some commonly used parameters and the overall experimental method we use for the simulated observed point clouds in our experiments in Subsection \ref{sec:experiment-setup-nulldist}. We then specify the simulation protocol in Subsection \ref{sim-prot}. The simulation protocol requires: (1) various null models, models without acyclic structure, (2) various alternative models, models with acyclic structure, and (3) various test statistics. In Section \ref{sec:experiments-results}
we discuss the performance of the simulations and provide power calculations.


\subsection{Generating the null distribution}
\label{sec:experiment-setup-nulldist}

There are three parameters we need to consider in generating our experimental data: $N$ is the number of points in the point cloud, $S$ is the number of point clouds we draw, and $R$ is the number of times we repeat the testing procedure. We will set $R=5$,
$N\in\{25,50,100,500\}$, and $S=50$
in our simulations.

The general simulation procedure for the null distribution can be summarized as follows, repeat $R$ times
\begin{enumerate}
    \item Draw $N$ points from the model to form an observed point cloud $X$
    \item Compute all topological invariants (see below for complete list) of interest
    \item For each estimation method out of  $\{$unbiased axis-aligned, convex hull, unbiased convex hull$\}$ estimate a null model
    \item Draw $S$ point clouds, each with $N$ points from the estimated null model
    \item Compute all topological invariants of interest of each of the $S$ simulated point clouds
\end{enumerate}

These computations provide us for each simulated observed point cloud $X$ with several draws from a corresponding empirical null distribution based on which we perform our testing procedures. The idea is to treat each topological invariant and each point cloud specification as an empirical distribution and draw at random with replacement from this empirical distribution.

We used the R re-implementation ripserr \cite{ripserr} of the ripser \cite{Bauer2021Ripser} Vietoris-Rips computation package for the topological calculations.

The computations are all performed in R (4.2.0) \cite{Rlang} using packages
tidyverse (1.3.2) \cite{tidyverse},
parallel \cite{Rlang},
doParallel (1.0.17) \cite{doParallel},
foreach (1.5.2) \cite{foreach},
mvtnorm (1.1) \cite{mvtnorm},
spatstat (2.4) \cite{spatstat},
volesti (1.1.2) \cite{volesti},
geometry (0.4.6.1) \cite{geometry},
ripserr (0.2.0) \cite{ripserr}.

\subsection{The simulation protocol}
\label{sim-prot}

\subsubsection{The null models -- acyclic models}

We will consider seven different null models.

\begin{enumerate}
\item{Axis-aligned rectangles} -- 
We draw $N$ points uniformly at random from the rectangle (box) with the side lengths either specified as 
\texttt{null.axis($w$,$h$)} or
\texttt{null.axis($w$,$h$,$d$)}
Here $w,h,d \in \{0.1,1,10\}$.

\item{Cross polytopes} -- 
For each combination of  $(x,y)$ or  $(x,y,z)$ values with $x,y,z \in \{0.1,1,10\}$, we draw $N$ points uniformly at random from the convex hull of the points $\{(\pm x,0), (0,\pm y)\}$ or  $\{(\pm x,0,0), (0,\pm y,0), (0,0,\pm z)\}$. These are denoted \texttt{null.cross($x$,$y$)} or \texttt{null.cross($x$,$y$,$z$)}.

\item{Random H-polytopes} -- 
An H-polytope is an intersection of half-spaces. We will draw $f$ $d$-dimensional halfspaces and generate
a $d$-dimensional H-polytope as the intersection of $f$ half-spaces. We sample $N$ points uniformly at random from the resulting H-polytopes. We 
set $d\in\{2,\dots,6\}$ and $f\in\{10,25,50,75,100\}$. We denote the
null models as \texttt{null.random.polytope($d$,$f$)}.

\item{Random V-polytopes}
A V-polytope is a convex hull of a collection of vertices. We will draw $v$
points randomly from the unit sphere in $\mathbb{R}^d$ and the V-polytope will be the convex hull of these points. We then sample $N$ points uniformly from the resulting V-polytope. We set $d\in\{2,\dots,6\}$ and $v\in\{10,25,50,75,100\}$. We denote the
null models as \texttt{null.random.hull($d$,$v$)}.

\item{Canonical Simplex} -- 
The $d$-dimensional canonical simplex is the convex hull of $d$ unit vectors in $\mathbb{R}^d$. We sample $N$ points uniformly at random from the canonical simplices. We set $d\in\{3,\dots,6\}$. We denote the null model as  \texttt{null.simplex(canonical,$d$)}.

\item{Unit simplex} --
The $d$-dimensional unit simplex is the cone with cone point $0$ and the base is
the $d$-dimensional canonical simplex. We sample $N$ points uniformly from the unit simplex. We set $d\in\{2,\dots,4\}$. We denote the null model as \texttt{null.simplex(unit,$d$)}.

\item{Unit Ball} --
We consider the unit ball in $\mathbb{R}^d$ and sample $N$ points uniformly at random. We set  $d\in\{2,\dots,5\}$. We denote the null model as \texttt{null.ball($d$)}.

\end{enumerate}

\subsubsection{The alternative -- cyclic models}

For the alternative models the support of the generative process will have structure that can be picked up by topological summaries of the data.

We consider four models for the alternative.

\begin{enumerate}
\item{Sphere} -- 
Sample $(x_1,...,x_N)$
uniformly from the unit sphere 
$\mathbb{S}^d$,
where $d=\{2,3\}$. 
To each $x_i$ add $\varepsilon_i \stackrel{iid}{\sim} \mbox{N}(0,\sigma^2 I_d)$
where $\sigma^2 = \{0.01,0.05,0.1\}$.
We denote this model as \texttt{power.sphere($d$).mvn.$\sigma^2$}.

\item{Concentric circles} -- 
For each outer radius $r\in\{1.25, 2, 5, 10\}$ we sample $N/(1+r)$ points from the circle with center 0 and radius $r$ and $N-N/(1+r)$ points from the unit circle. 
To each $x_i$ add $\varepsilon_i \stackrel{iid}{\sim} \mbox{N}(0,\sigma^2 I_d)$
where $\sigma^2 = \{0.01,0.05,0.1\}$.
We denote this model as
\texttt{power.concentric($r$).mvn.$\sigma^2$}.

\item{Figure 8} -- 
For each radius $r\in\{0.25,0.5,1.0,1.5,5.0\}$ two osculating circles are generated: the unit circle with center $(1,0)$ and the circle with radius $r$ and center $(-r,0)$. We draw $N/(1+r)$ points uniformly at random from the circle with radius $r$ and $N-N/(1+r)$ points from the unit circle. To each $x_i$ add $\varepsilon_i \stackrel{iid}{\sim} \mbox{N}(0,\sigma^2 I_d)$
where $\sigma^2 = \{0.01,0.05,0.1\}$.
We denote this model as \texttt{power.fig8($r$).mvn.$\sigma^2$}.

\item{Thomas cluster process} --
The Thomas process can be stated as a hierarchical model\\
\indent 1) Draw the number of cluster points -- $p \sim \mbox{Pois}(\kappa)$; \\
\indent 2) Draw the cluster points -- $(x_1,...,x_p)  \stackrel{iid}{\sim}\mbox{U}([0,1]^d)$ ; \\
\indent 3) Draw the number of points in each cluster 
$n_i \sim \mbox{Pois}(\mu)$ for $i=1,...,p$
\\
\indent 4) Draw the points from the clustered Thomas process -- $(d_{i1},...,d_{in_i}) \stackrel{iid}{\sim} \mbox{N}(x_i,\sigma^2 \mbox{Id})$ for $i=1,...,p$
The parameters are the intensity $\kappa$, the scale $\sigma$, and the expected cluster size $\mu$.
The expected number of points in each draw is $\kappa\cdot\mu$.
For each scale $\sigma\in\{0.05,0.1,0.15,0.2,0.25\}$ we draw from a Thomas process with $\kappa=5$ and $\mu=N/5$. We denote this model as
 \texttt{power.thomas($\sigma$)}.

\end{enumerate}


For the Sphere with $d=2$ we show some examples of these point clouds in Figure \ref{fig:power-sphere}. For the concentric circle
with $d=2$ we show some examples of these point clouds in Figure \ref{fig:power-concentric}. For the other model types we show corresponding figures in Appendix \ref{app:point-clouds}.
For the figure 8 we show some examples in
Figure \ref{fig:power-fig8}
We show some examples of the point clouds generated by the Thomas process 
in Figure \ref{fig:power-thomas}.

\begin{figure}
    \centering
\foreach\ss in {0.01,0.05,0.1} {
$\sigma^2=\ss$ \\
  \foreach\rr in {2} {
    \foreach\NN in {25,50,100,500} {
      \includegraphics[height=1.75cm,trim=0 0 550 0,clip]{pointcloud-figures/power.sphere-\rr-.mvn.\ss-\NN.axis.pdf}
    }
    \\\vspace{-3ex}
  }
}
    \caption{Circle point clouds for a range of noise levels ($\sigma^2$) and point cloud sizes (25, 50, 100 and 500 points).}
    \label{fig:power-sphere}
\end{figure}

\subsection{Null estimation models}

Based on the null model estimation methods we discussed in \ref{sec:null-distributions-one-sample-test}, we identify three potentially interesting ways to estimate a null distribution and drawing from it.

\begin{figure}
    \centering
  \includegraphics[height=1.75cm]{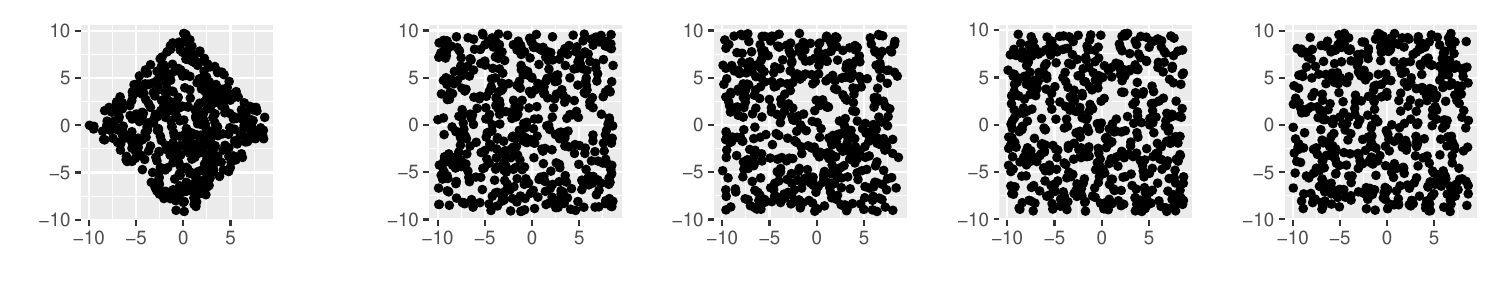} \\
    \includegraphics[height=1.75cm]{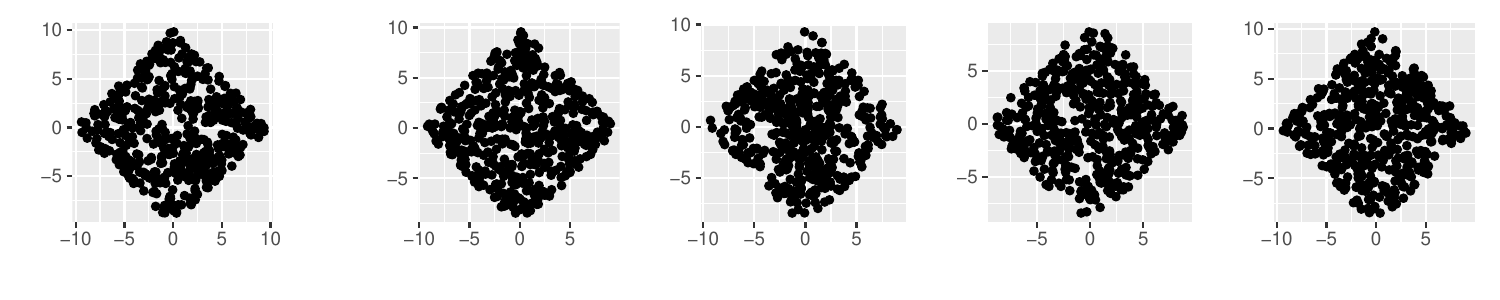} \\
    \includegraphics[height=1.75cm]{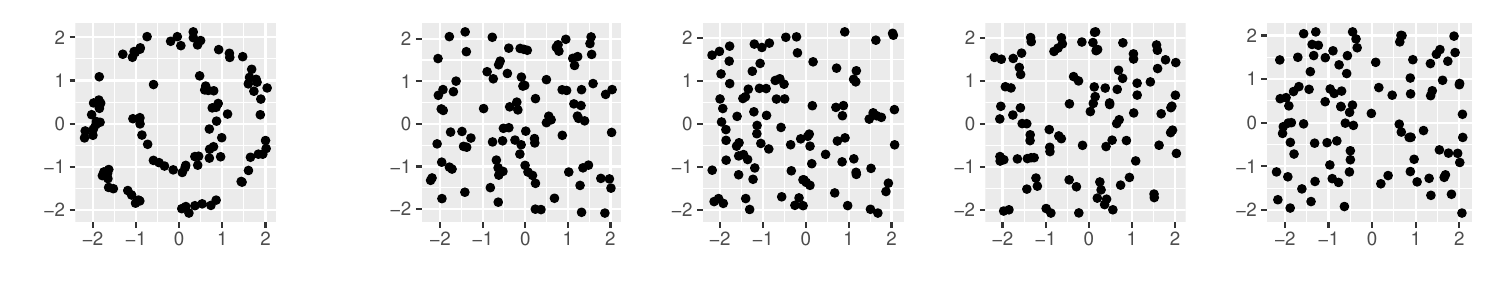} \\
  \includegraphics[height=1.75cm]{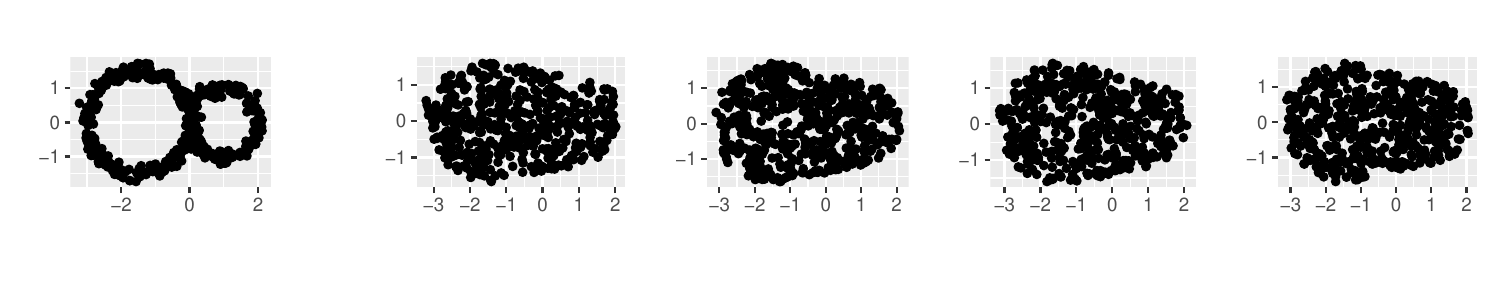} \\
  \includegraphics[height=1.75cm]{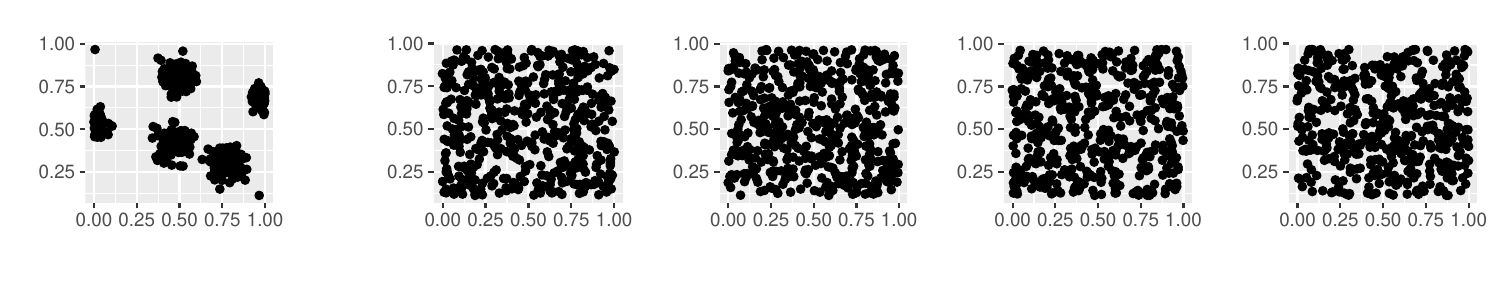} \\
        \caption{Some examples of point clouds and their axis-aligned or convex hull estimated null model point clouds. The first two rows are the cross-polytope acyclic model, third row is the concentric circles model with outer radius 2, fourth row is the figure 8 model with left radius 1.5 and the last row is a Thomas process with scale 0.05. \\
        The left-most column is the observed acyclic or cyclic model, and the four columns to the right are sample point clouds drawn from the estimated null model (using estimators, from the top, axis, hull, axis, hull, axis}.
    \label{fig:estimator}
\end{figure}

We show some examples of the first and second estimation methods in action in Figure \ref{fig:estimator}.

\subsubsection{Axis-aligned rectangles}
For each coordinate $r$ of the data $x=\{x^1,\dots,x^n\}$, estimate lower and upper bounds $(\hat a_r,\hat b_r)$ for a uniform distribution using the formulas
\[
{\hat a}_r = \frac{n\min_j\{x_r^j\} - \max_j\{x_r^j\}}{n-1}
\qquad
{\hat b}_r = \frac{n\max_j\{x_r^j\} - \min_j\{x_r^j\}}{n-1}
\]

To sample, draw uniformly at random each coordinate $x_r\sim\text{Uniform}(\hat a_r, \hat b_r)$ to assemble the next observation $x=(x_1,\dots,x_d)$.

\subsubsection{Convex hull}
Compute the convex hull $H(x)$ of the data $x$. Sample by drawing uniformly at random from this convex hull.

\subsubsection{Unbiased convex hull}
Compute the convex hull $H(x)$ and its set of boundary vertices $V_\partial$. Since the centroid of the convex hull is complicated to compute, we approximate it with the centroid $c_\partial$ of the boundary vertices of $H(x)$. Compute
\[
\hat{H}(x) = \sqrt{\frac{|x|}{|x|-|V_\partial|}}(H(x) - c_\partial)
\]
Up to translation, this is an unbiased estimate of the convex hull.

To sample, draw points uniformly at random from $\hat{H}(x)$.

\subsection{Topological invariants}

There is a wide range of numeric invariants of persistence diagrams in active use in the community. The classically most commonly used is the ``length of the longest bar'' (which one might call $L_\infty$ persistence norm). Also commonly used are $L_1$ and $L_2$ versions, taking the ``mean bar length'' or ``root mean bar length square''.

In addition to these ``additive'' invariants, \cite{bobskra} have studied the universality of distributions of invariants deriving from a ``multiplicative'' or ``log-scale'' persistence invariant. For a persistence bar $(b,d)$ with birth parameter value $b$ and death parameter value $d$, instead of the $d-b$ used in the additive persistence norms, Bobrowski and Skraba study invariants defined in terms of $\log\log(d/b)$. The result is invariants that measure shape independent of not just translation, but independent also of scaling of point clouds.

\cite{bobskra} define two families of invariants, defined for each persistence bar separately:
\begin{itemize} 
    \item $\pi=d/b$.
    \item $\ell=\log\log(d/b) - \mu -\gamma$, where $\mu$ is the mean of all the $\log\log(d/b)$ quantities, and $\gamma\approx0.5772156649$ is the Euler-Mascheroni constant.
\end{itemize}
These are subsequently summarized across all the persistence bars. We use the $L_1, L_2, L_\infty$ norms as for the additive invariants to summarize $\pi$ and $\ell$.

We use 21 invariants in our data generation:

For each homological dimension $d=\{0,1,2\}$, and for each of $L_1,L_2,L_\infty$ we compute the corresponding persistence norm of the Betti $d$ persistence diagram.
We denote these as \texttt{L1.$d$}, \texttt{L2.$d$} and \texttt{Linf.$d$}.

For each homological dimension $d=\{1,2\}$, and each of $L_1,L_2,L_\infty$, and each of $\pi, \ell$ we compute the corresponding norm of the corresponding bar invariant of the Betti $d$ persistence diagram.
We denote these as \texttt{pi.L$p$.$d$} and \texttt{ell.L$p$.$d$}, where $p\in\{1,2,\infty\}$.

\section{Experiments: Questions and results}
\label{sec:experiments-results}

There are a number of questions we intend to answer by using these pre-computed topological invariants.

\begin{itemize}
    \item Can these standardized topological invariants really be compared with each other?
    \item Validate and measure statistical power for the one-sample test described in Method \ref{mth:one-sample-simulation}.
    \item Evaluate whether the unbiased convex hull estimator is really needed.
    \item Validate and measure power for the FWER correction method described in Method \ref{mth:fwer}.
    \item Validate and measure power for a composite FWER correction method, using several different invariants. A natural approach is to combine all homological dimensions into one composite test.
    \item Evaluate the differences in performance between the additive and the two multiplicative invariants.
\end{itemize}

For our validation attempts, we shall in each case study both how close the p-value distribution is to uniform, and what the corresponding empirical levels are for cutoffs chosen to produce $\alpha\in\{0.01,0.05,0.10\}$.

We shall proceed to address each of these questions with corresponding computations now.

\subsection{Exchangeability}
\label{sec:exchangeability}

Can our invariants when computed on point clouds of different sizes and shapes really be compared with each other? We have reasons to believe so -- as described in Section \ref{sec:comparing-statistics}, as well as in the work by Bobrowski and Skraba \cite{bobskra}.

\begin{figure}
    \centering
    \includegraphics[width=\linewidth]{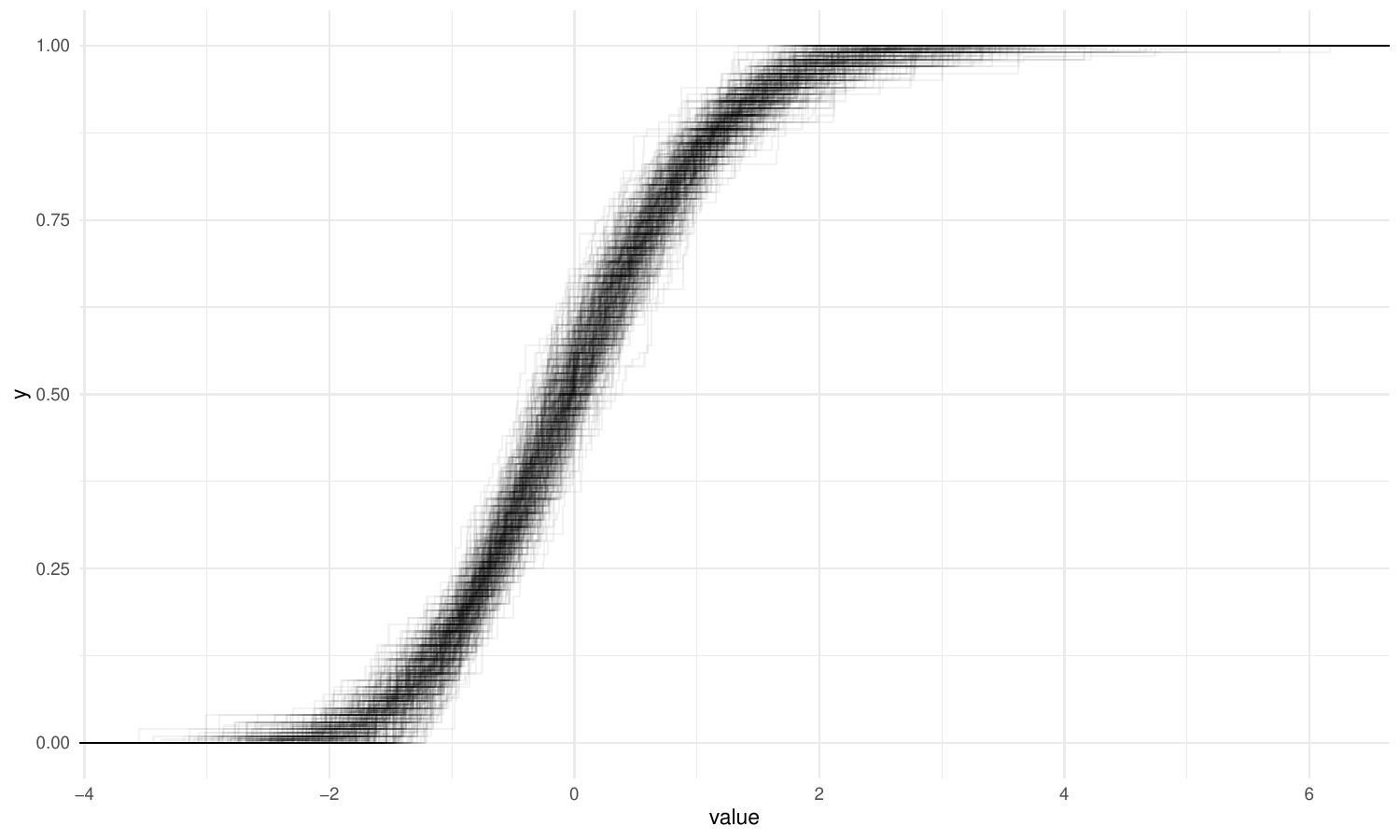}
    \caption{ECDFs of each topological invariant of a sample of 100 standardized simulated point clouds for each of 100 different randomly selected acyclic model setups.}
    \label{fig:exchangeability-ecdf}
\end{figure}

For an experimental investigation, we draw 100 acyclic models at random from our full collection of models, estimate and simulate 100 point clouds for each model, and standardize all topological invariants for each of these models. In Figure \ref{fig:exchangeability-ecdf} we show the ECDF for all of these options and all the invariants in one single plot. While there is some variation around a core curve, notice there are no dramatic outliers among all these curves.

\subsection{Bias of the convex hull}

Consider the axis-aligned rectangles. We know that estimating these with axis-aligned rectangles is going to give us an unbiased result.

We repeat 100 times drawing an axis-aligned rectangle acyclic model at random from the 16 different versions we have defined. For each acyclic model, we estimate it in three different ways, and for each resulting null model we draw 100 null model point clouds. We compute the p-values, and extract 100 different p-values for each topological invariant and each estimation method.

\begin{figure}
    \centering
    \includegraphics[width=\linewidth]{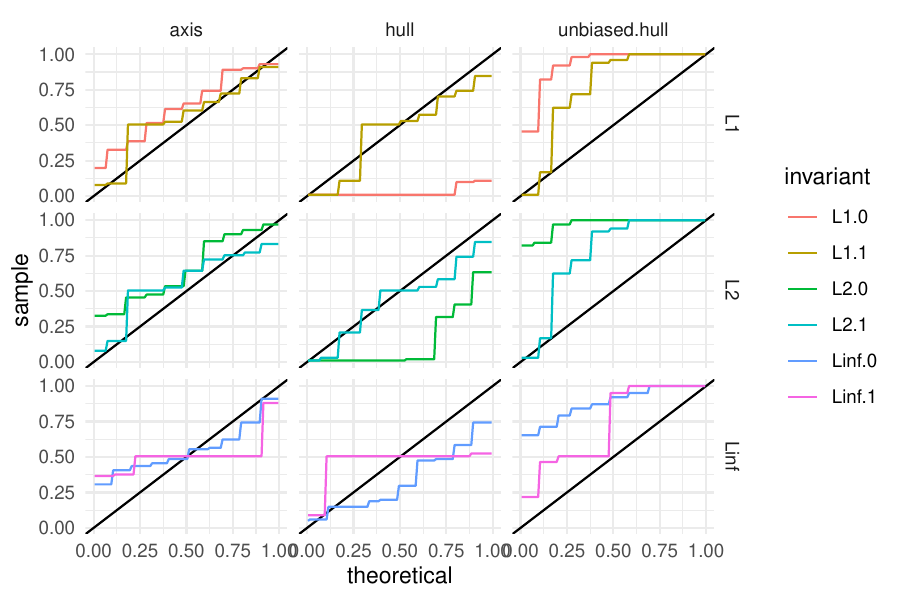}
    \caption{In the left-most column, what we would expect accurate and unbiased estimation and null models to look like. In the middle, the undilated convex hull results. To the right, the results from dilating the convex hull estimation as described. Best results among the convex hull approaches are achieved with the $L_\infty$ persistence norm methods and the dilated unbiased convex hull estimation method, while $L_1$ and $L_2$ perform worse in the unbiased convex hull approach, and all the biased hull methods perform worse again.}
    \label{fig:qq-axis-null-validation}
\end{figure}

Since these are p-values from a null-model, we should expect them to be uniformly distributed. In Figure \ref{fig:qq-axis-null-validation}, we see the QQ-plots against a uniform distribution split up by estimator and by persistence norm.

The consequences of these discrepancies becomes clear when we look in Table \ref{tab:rejection-axis-acyclic} at rejection rates at different intended rejection levels $\alpha$: the rejection rates for the biased convex hull column -- especially in homological dimension 0 -- look more like power computation numbers than level estimations. The bias is large enough to skew the additive topological invariants enough to give massive amounts of false positives (94\% false positives at $\alpha=0.01$ if one were using \texttt{L1.0}).

\begin{table}
    \centering
    \begin{tabular}{lccc}
    \toprule
        Invariant & axis & hull & unbiased hull  \\ \midrule
        \texttt{Linf.0} & .08/.15/.28 & .21/.29/.38 & .02/.05/.05 \\
        \texttt{Linf.1} & .03/.07/.10 & .02/.10/.22 & .03/.03/.05 \\ \midrule
        \texttt{L1.0} & .00/.00/.08 & .94/.96/.96 & .00/.03/.09 \\
        \texttt{L1.1} & .06/.06/.24 & .10/.26/.36 & .08/.08/.10 \\ \midrule
        \texttt{L2.0} & .05/.09/.13 & .82/.88/.92 & .00/.00/.06 \\
        \texttt{L2.1} & .06/.13/.28 & .22/.32/.44 & .05/.08/.12 \\ \midrule\bottomrule
    \end{tabular}
    \caption{Rejection rates by estimator and invariant for an axis-aligned rectangular acyclic model. Rates for $\alpha=0.01/0.05/0.10$ are separated with $/$ in each table cell.}
    \label{tab:rejection-axis-acyclic}
\end{table}

In subsequent experiments we shall ignore the biased convex hull estimator.

\subsection{Validate the one-hypothesis one-sample test}

For each of the acyclic model types, we repeat 100 times a drawing one acyclic model at random, and then estimate it with the axis and unbiased hull methods and draw 100 null model point clouds. We compute p-values and extract the resulting 100 different p-values for each topological invariant and estimation method.

\begin{figure}
    \centering
    \includegraphics[width=\linewidth]{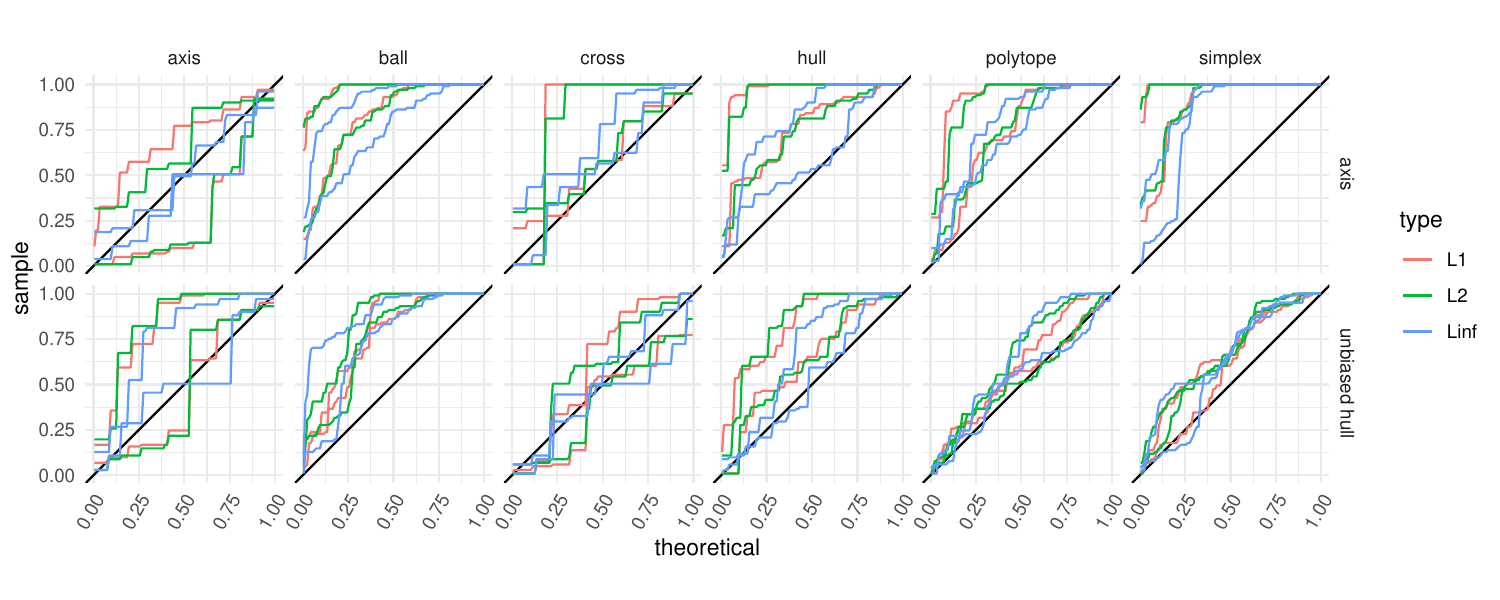}
    \caption{QQ-plot for p-values split up by acyclic model type and by estimator type. The fact that these QQ-curves almost all lie over the diagonal line means that in almost all cases, we under-produce false positives.}
    \label{fig:qq-acyclic-models-validation}
\end{figure}

In Figure \ref{fig:qq-acyclic-models-validation} we see the QQ-plots for each type of acyclic model we use and each type of additive invariant. Most cases will end up under-producing false positives according to this plot, something can see reflected in Table \ref{tab:rejection-by-acyclic}.

\begin{table}
    \centering
    \begin{tabular}{llcc}
    \toprule
        Model & Invariant & axis & unbiased hull  \\ \midrule
\multirow{6}{*}{axis}
& \texttt{L1.0} & 0.00/0.00/0.00 & 0.00/0.00/0.00 \\ 
& \texttt{L1.1} & 0.11/0.21/0.55 & 0.00/0.00/0.19 \\ 
& \texttt{L2.0} & 0.00/0.00/0.00 & 0.00/0.00/0.00 \\ 
& \texttt{L2.1} & 0.21/0.31/0.42 & 0.00/0.08/0.15 \\ 
& \texttt{Linf.0} & 0.00/0.00/0.00 & 0.00/0.00/0.00 \\ 
& \texttt{Linf.1} & 0.00/0.10/0.10 & 0.00/0.08/0.08 \\ 
\midrule\multirow{6}{*}{ball}
& \texttt{L1.0} & 0.00/0.00/0.00 & 0.00/0.02/0.03 \\ 
& \texttt{L1.1} & 0.00/0.00/0.00 & 0.00/0.00/0.00 \\ 
& \texttt{L2.0} & 0.00/0.00/0.00 & 0.00/0.01/0.01 \\ 
& \texttt{L2.1} & 0.00/0.00/0.00 & 0.00/0.00/0.00 \\ 
& \texttt{Linf.0} & 0.00/0.00/0.00 & 0.00/0.00/0.01 \\ 
& \texttt{Linf.1} & 0.00/0.02/0.02 & 0.01/0.01/0.01 \\ 
\midrule\multirow{6}{*}{cross}
& \texttt{L1.0} & 0.18/0.18/0.18 & 0.13/0.13/0.21 \\ 
& \texttt{L1.1} & 0.00/0.00/0.00 & 0.00/0.22/0.32 \\ 
& \texttt{L2.0} & 0.18/0.18/0.18 & 0.13/0.13/0.21 \\ 
& \texttt{L2.1} & 0.00/0.00/0.00 & 0.00/0.00/0.32 \\ 
& \texttt{Linf.0} & 0.11/0.11/0.19 & 0.13/0.13/0.23 \\ 
& \texttt{Linf.1} & 0.00/0.00/0.00 & 0.00/0.00/0.11 \\ 
\midrule\multirow{6}{*}{random.hull}
& \texttt{L1.0} & 0.00/0.00/0.00 & 0.00/0.00/0.00 \\ 
& \texttt{L1.1} & 0.00/0.00/0.00 & 0.10/0.10/0.10 \\ 
& \texttt{L2.0} & 0.00/0.00/0.00 & 0.00/0.00/0.00 \\ 
& \texttt{L2.1} & 0.00/0.00/0.00 & 0.10/0.10/0.10 \\ 
& \texttt{Linf.0} & 0.00/0.00/0.02 & 0.00/0.00/0.09 \\ 
& \texttt{Linf.1} & 0.00/0.02/0.02 & 0.01/0.04/0.09 \\ 
\midrule\multirow{6}{*}{random.polytope}
& \texttt{L1.0} & 0.00/0.00/0.00 & 0.02/0.02/0.07 \\ 
& \texttt{L1.1} & 0.00/0.02/0.07 & 0.01/0.03/0.04 \\ 
& \texttt{L2.0} & 0.00/0.00/0.00 & 0.00/0.02/0.07 \\ 
& \texttt{L2.1} & 0.00/0.02/0.04 & 0.01/0.03/0.08 \\ 
& \texttt{Linf.0} & 0.00/0.05/0.05 & 0.04/0.04/0.04 \\ 
& \texttt{Linf.1} & 0.00/0.00/0.05 & 0.00/0.02/0.13 \\ 
\midrule\multirow{6}{*}{simplex}
& \texttt{L1.0} & 0.00/0.00/0.00 & 0.01/0.03/0.09 \\ 
& \texttt{L1.1} & 0.00/0.00/0.00 & 0.00/0.02/0.04 \\ 
& \texttt{L2.0} & 0.00/0.00/0.00 & 0.02/0.03/0.05 \\ 
& \texttt{L2.1} & 0.00/0.00/0.00 & 0.00/0.00/0.02 \\ 
& \texttt{Linf.0} & 0.01/0.02/0.02 & 0.00/0.03/0.11 \\ 
& \texttt{Linf.1} & 0.00/0.00/0.00 & 0.00/0.02/0.03 \\ 
\midrule
\\\bottomrule
\end{tabular}
    \caption{Rejection rates by acyclic model, estimator and invariant for each type of acyclic model. Rates for $\alpha=0.01/0.05/0.10$ are separated with $/$ in each table cell.}
    \label{tab:rejection-by-acyclic}
\end{table}

If we look instead at overall rejection rates, without splitting them up at all, we get the results in Table \ref{tab:rejection-all-acyclic}. These rejection rates are not terribly far off from the intended 0.01/0.05/0.10.

\begin{table}
    \centering
    \begin{tabular}{llcc}
    \toprule
        Invariant & axis & unbiased hull  \\ \midrule
\texttt{L1.0} & 0.06/0.07/0.08 & 0.09/0.10/0.15 \\ 
\texttt{L1.1} & 0.03/0.05/0.07 & 0.02/0.03/0.06 \\ 
\texttt{L2.0} & 0.08/0.08/0.10 & 0.05/0.11/0.14 \\ 
\texttt{L2.1} & 0.06/0.07/0.09 & 0.02/0.07/0.07 \\ 
\texttt{Linf.0} & 0.07/0.10/0.11 & 0.01/0.03/0.08 \\ 
\texttt{Linf.1} & 0.00/0.02/0.07 & 0.00/0.02/0.04 \\ 

\\\bottomrule
    \end{tabular}
    \caption{Rejection rates across all acyclic models. Rates for $\alpha=0.01/0.05/0.10$ are separated with $/$ in each table cell.}
    \label{tab:rejection-all-acyclic}
\end{table}

\subsection{Power of the one-sample test}

We expect the power of the on-sample test to vary with the sample size, the topological shape to be discovered, the null estimator, and the amount of noise in the data. To estimate this power, we sample 100 times with replacement from the 5 iterations of data computed for each combination of noise levels, sample sizes, dimensions, estimators. For each sampled iteration, we compute 100 simulated point clouds and use Method \ref{mth:one-sample-simulation} to generate a p-value. This p-value is used against levels of 0.01, 0.05 and 0.10 to generate rejections.

In Table \ref{tab:power-one-sample}, we show the power of the one-sample test to detect a topological structure, separated by noise levels, sample sizes, estimators, and invariants used.

We include ECDF-plots for each type of power estimation model in Appendix \ref{app:power-plots}.

\begin{table}
    \centering
    \begin{tabular}{llccccc}
    \toprule
        Estimator & Invariant & Noise & 25 & 50 & 100 & 500  \\ \midrule
\multirow{9}{*}{axis} & \multirow{3}{*}{L1} & 0.01 & 0.80/0.87/0.97  & 0.80/0.80/0.80  & 1.00/1.00/1.00  & 0.00/0.00/0.00 \\ 
 & & 0.05 & 0.20/0.20/0.20  & 0.20/0.27/0.38  & 0.00/0.00/0.00  & 0.00/0.00/0.00 \\ 
 & & 0.1 & 0.20/0.35/0.40  & 0.00/0.12/0.20  & 0.00/0.00/0.00  & 0.00/0.00/0.00 \\ 
 \cmidrule(lr){2-7} & \multirow{3}{*}{L2} & 0.01 & 0.82/0.99/1.00  & 0.80/0.80/0.80  & 1.00/1.00/1.00  & 1.00/1.00/1.00 \\ 
 & & 0.05 & 0.20/0.20/0.21  & 0.60/0.60/0.60  & 0.41/0.53/0.71  & 0.80/0.80/0.80 \\ 
 & & 0.1 & 0.20/0.35/0.40  & 0.02/0.19/0.25  & 0.00/0.00/0.00  & 0.00/0.00/0.00 \\ 
 \cmidrule(lr){2-7} & \multirow{3}{*}{Linf} & 0.01 & 0.26/0.60/0.62  & 0.80/0.80/0.80  & 1.00/1.00/1.00  & 1.00/1.00/1.00 \\ 
 & & 0.05 & 0.02/0.02/0.02  & 0.43/0.59/0.60  & 0.49/0.98/1.00  & 1.00/1.00/1.00 \\ 
 & & 0.1 & 0.40/0.40/0.40  & 0.00/0.27/0.60  & 0.00/0.16/0.37  & 0.40/0.41/0.48 \\ 
 \cmidrule(lr){2-7}\midrule
\multirow{9}{*}{unbiased.hull} & \multirow{3}{*}{L1} & 0.01 & 0.62/0.79/0.80  & 0.60/0.74/0.80  & 0.83/0.99/1.00  & 0.00/0.00/0.00 \\ 
 & & 0.05 & 0.20/0.20/0.22  & 0.40/0.46/0.59  & 0.00/0.05/0.19  & 0.00/0.00/0.00 \\ 
 & & 0.1 & 0.20/0.23/0.35  & 0.00/0.12/0.20  & 0.00/0.00/0.00  & 0.00/0.00/0.00 \\ 
 \cmidrule(lr){2-7} & \multirow{3}{*}{L2} & 0.01 & 0.80/0.80/0.91  & 0.80/0.80/0.80  & 1.00/1.00/1.00  & 1.00/1.00/1.00 \\ 
 & & 0.05 & 0.20/0.20/0.30  & 0.60/0.60/0.60  & 0.63/0.78/0.80  & 1.00/1.00/1.00 \\ 
 & & 0.1 & 0.20/0.26/0.38  & 0.40/0.40/0.40  & 0.00/0.00/0.10  & 0.00/0.00/0.00 \\ 
 \cmidrule(lr){2-7} & \multirow{3}{*}{Linf} & 0.01 & 0.40/0.47/0.58  & 0.80/0.80/0.80  & 1.00/1.00/1.00  & 1.00/1.00/1.00 \\ 
 & & 0.05 & 0.20/0.23/0.35  & 0.60/0.60/0.64  & 1.00/1.00/1.00  & 1.00/1.00/1.00 \\ 
 & & 0.1 & 0.06/0.21/0.22  & 0.24/0.52/0.66  & 0.40/0.56/0.63  & 0.42/0.60/0.66 \\ 
 \cmidrule(lr){2-7}
\\\bottomrule
 \end{tabular}
    \caption{Rejection rates across all Figure 8 power models, using homological dimension 1. Rates for $\alpha=0.01/0.05/0.10$ are separated with $/$ in each table cell.}
    \label{tab:power-one-sample}
\end{table}

\subsection{Bobrowski-Skraba multiplicative invariants}

The multiplicative invariants $\pi$ and $\ell$ from \cite{bobskra} are designed to be scale-invariant as well as translation-invariant. This reflects in our setting in these invariants handling the unbiased convex hull far better than the additive invariants.

To validate our one-sample hypothesis test with multiplicative invariants, we perform 100 experiments for each acyclic model type; in each experiment we draw one acyclic model at random, and estimate it with all three estimators and draw 100 null model point clouds. We compute p-values and extract the resulting 100 different p-values for each multiplicative invariant and estimation method. The resulting rejection rates at levels of $\alpha=0.01, 0.05, 0.10$ are reported in Table \ref{tab:rejection-by-acyclic-multiplicative}.

\begin{table}
    \centering
    \begin{tabular}{llccc}
    \toprule
        Model & Invariant & axis & hull & unbiased hull  \\ \midrule
\multirow{6}{*}{axis}& \texttt{ell.L1.1} & 0.00/0.00/0.00 & 0.00/0.00/0.00 & 0.00/0.00/0.00 \\ 
& \texttt{ell.L2.1} & 0.21/0.33/0.38 & 0.13/0.13/0.28 & 0.11/0.23/0.32 \\ 
& \texttt{ell.Linf.1} & 0.00/0.00/0.00 & 0.13/0.13/0.13 & 0.00/0.07/0.07 \\ 
& \texttt{pi.L1.1} & 0.11/0.23/0.56 & 0.26/0.34/0.42 & 0.11/0.26/0.41 \\ 
& \texttt{pi.L2.1} & 0.21/0.23/0.43 & 0.26/0.34/0.34 & 0.11/0.26/0.41 \\ 
& \texttt{pi.Linf.1} & 0.00/0.00/0.00 & 0.13/0.13/0.13 & 0.07/0.07/0.07 \\ 
\midrule\multirow{6}{*}{ball}& \texttt{ell.L1.1} & 0.00/0.00/0.00 & 0.00/0.00/0.00 & 0.00/0.00/0.00 \\ 
& \texttt{ell.L2.1} & 0.02/0.06/0.07 & 0.03/0.08/0.11 & 0.01/0.05/0.06 \\ 
& \texttt{ell.Linf.1} & 0.00/0.00/0.08 & 0.02/0.02/0.09 & 0.00/0.03/0.04 \\ 
& \texttt{pi.L1.1} & 0.01/0.03/0.06 & 0.00/0.02/0.07 & 0.00/0.09/0.18 \\ 
& \texttt{pi.L2.1} & 0.01/0.03/0.05 & 0.00/0.01/0.07 & 0.00/0.05/0.13 \\ 
& \texttt{pi.Linf.1} & 0.00/0.02/0.09 & 0.00/0.02/0.07 & 0.01/0.06/0.08 \\ 
\midrule\multirow{6}{*}{cross}& \texttt{ell.L1.1} & 0.00/0.00/0.00 & 0.00/0.00/0.00 & 0.00/0.00/0.00 \\ 
& \texttt{ell.L2.1} & 0.00/0.00/0.00 & 0.00/0.00/0.00 & 0.00/0.00/0.00 \\ 
& \texttt{ell.Linf.1} & 0.00/0.00/0.00 & 0.00/0.00/0.00 & 0.00/0.00/0.00 \\ 
& \texttt{pi.L1.1} & 0.18/0.31/0.31 & 0.12/0.33/0.33 & 0.11/0.22/0.31 \\ 
& \texttt{pi.L2.1} & 0.18/0.18/0.31 & 0.12/0.24/0.36 & 0.11/0.22/0.22 \\ 
& \texttt{pi.Linf.1} & 0.00/0.00/0.00 & 0.00/0.00/0.12 & 0.00/0.00/0.21 \\ 
\midrule\multirow{6}{*}{random.hull}& \texttt{ell.L1.1} & 0.00/0.00/0.00 & 0.00/0.00/0.00 & 0.00/0.00/0.00 \\ 
& \texttt{ell.L2.1} & 0.00/0.02/0.11 & 0.02/0.11/0.11 & 0.03/0.03/0.06 \\ 
& \texttt{ell.Linf.1} & 0.00/0.10/0.15 & 0.00/0.01/0.01 & 0.00/0.03/0.24 \\ 
& \texttt{pi.L1.1} & 0.16/0.16/0.16 & 0.18/0.19/0.20 & 0.14/0.14/0.14 \\ 
& \texttt{pi.L2.1} & 0.16/0.16/0.16 & 0.18/0.19/0.20 & 0.14/0.14/0.14 \\ 
& \texttt{pi.Linf.1} & 0.11/0.19/0.20 & 0.04/0.10/0.21 & 0.09/0.15/0.20 \\ 
\midrule\multirow{6}{*}{random.polytope}& \texttt{ell.L1.1} & 0.00/0.00/0.00 & 0.00/0.00/0.00 & 0.00/0.00/0.00 \\ 
& \texttt{ell.L2.1} & 0.03/0.09/0.15 & 0.02/0.05/0.12 & 0.01/0.01/0.08 \\ 
& \texttt{ell.Linf.1} & 0.00/0.00/0.04 & 0.00/0.00/0.06 & 0.00/0.00/0.02 \\ 
& \texttt{pi.L1.1} & 0.01/0.09/0.13 & 0.07/0.09/0.11 & 0.02/0.06/0.11 \\ 
& \texttt{pi.L2.1} & 0.01/0.09/0.15 & 0.04/0.06/0.11 & 0.01/0.07/0.11 \\ 
& \texttt{pi.Linf.1} & 0.00/0.02/0.02 & 0.00/0.00/0.08 & 0.00/0.06/0.11 \\ 
\midrule\multirow{6}{*}{simplex}& \texttt{ell.L1.1} & 0.00/0.00/0.00 & 0.00/0.00/0.00 & 0.00/0.00/0.00 \\ 
& \texttt{ell.L2.1} & 0.01/0.01/0.07 & 0.01/0.03/0.09 & 0.05/0.07/0.14 \\ 
& \texttt{ell.Linf.1} & 0.01/0.01/0.05 & 0.02/0.02/0.02 & 0.00/0.04/0.09 \\ 
& \texttt{pi.L1.1} & 0.02/0.03/0.05 & 0.01/0.02/0.06 & 0.00/0.05/0.07 \\ 
& \texttt{pi.L2.1} & 0.01/0.02/0.07 & 0.01/0.02/0.07 & 0.00/0.05/0.07 \\ 
& \texttt{pi.Linf.1} & 0.01/0.03/0.03 & 0.00/0.00/0.02 & 0.01/0.05/0.05 \\ 
\midrule
\\\bottomrule
    \end{tabular}
    \caption{Rejection rates by acyclic model, estimator and invariant for each type of acyclic model using the multiplicative invariants $\pi$ and $\ell$. Rates for $\alpha=0.01/0.05/0.10$ are separated with $/$ in each table cell.}
    \label{tab:rejection-by-acyclic-multiplicative}
\end{table}

The overall rejection rates (not splitting by acyclic model type) computed by sampling 100 acyclic models from each type, but summarizing the rejection rates without splitting the count on model types are reported in Table \ref{tab:rejection-all-acyclic-multiplicative}.

\begin{table}
    \centering
    \begin{tabular}{llccc}
    \toprule
        Invariant & axis & hull & unbiased hull  \\ \midrule
\texttt{ell.L1.1} & 0.00/0.00/0.00 & 0.00/0.00/0.00 & 0.00/0.00/0.00 \\ 
\texttt{ell.L2.1} & 0.04/0.09/0.13 & 0.04/0.07/0.12 & 0.04/0.07/0.11 \\ 
\texttt{ell.Linf.1} & 0.00/0.02/0.05 & 0.03/0.03/0.05 & 0.00/0.03/0.08 \\ 
\texttt{pi.L1.1} & 0.08/0.14/0.21 & 0.11/0.17/0.20 & 0.06/0.14/0.20 \\ 
\texttt{pi.L2.1} & 0.10/0.12/0.20 & 0.10/0.14/0.19 & 0.06/0.13/0.18 \\ 
\texttt{pi.Linf.1} & 0.02/0.04/0.06 & 0.03/0.04/0.10 & 0.03/0.07/0.12 \\ 

\\\bottomrule
\end{tabular}
    \caption{Rejection rates across all types of acyclic model, for each estimator and invariant using the multiplicative invariants $\pi$ and $\ell$. Rates for $\alpha=0.01/0.05/0.10$ are separated with $/$ in each table cell.}
    \label{tab:rejection-all-acyclic-multiplicative}
\end{table}

\subsection{Validate the FWER control method}

To validate the FWER control method, we do the following experiment 100 times:

\begin{enumerate}
    \item Draw a number $k$ of observed point clouds at random from a Poisson distribution with mean 10.
    \item Draw $k$ acyclic models at random from all the models we have defined.
    \item For each estimator (axis and unbiased hull) draw 100 null model point clouds for each observed point cloud, and compute standardized topological invariants for each.
    \item Maximize invariants, rank and compute p-values.
\end{enumerate}

\begin{figure}
    \centering
    \includegraphics[width=\linewidth]{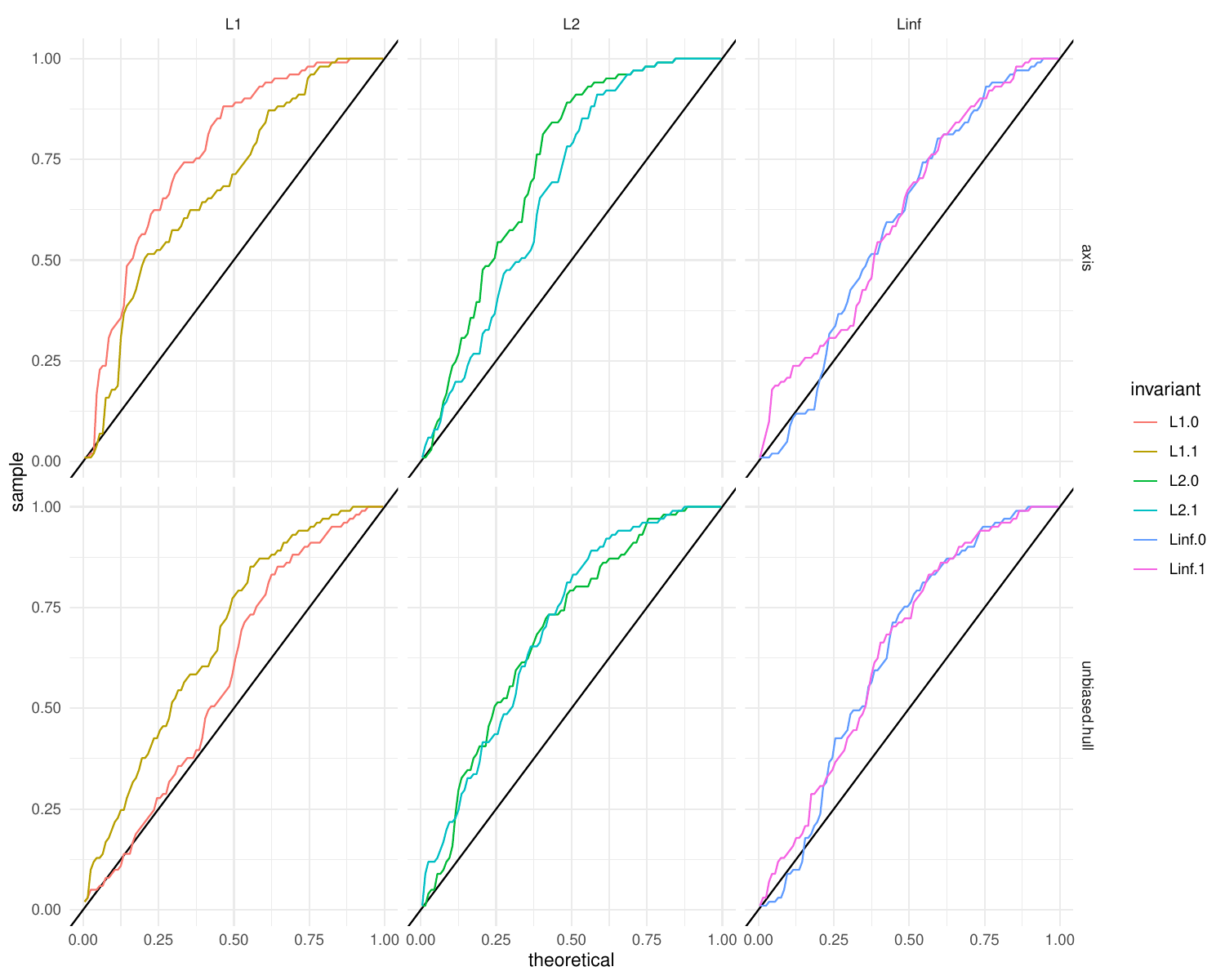}
    \caption{QQ-plots for the FWER procedure on collections of acyclic point clouds against a uniform distribution.}
    \label{fig:qq-fwer-validation}
\end{figure}

The resulting p-values are displayed in QQ-plots against a uniform distribution in Figure \ref{fig:qq-fwer-validation}, and corresponding rejection rates for $\alpha=0.01/0.05/0.10$ are reported in Table \ref{tab:rejection-fwer-acyclic}.

\begin{table}
    \centering
    \begin{tabular}{lcc}
    \toprule
        Invariant & axis & unbiased hull  \\ \midrule
ell.L1.1 & 0.00/0.00/0.00  & 0.00/0.00/0.00 \\ 
ell.L2.1 & 0.00/0.05/0.12  & 0.03/0.12/0.17 \\ 
ell.Linf.1 & 0.05/0.09/0.15  & 0.08/0.12/0.16 \\ 
L1.0 & 0.02/0.04/0.04  & 0.00/0.05/0.12 \\ 
L1.1 & 0.03/0.05/0.07  & 0.00/0.02/0.03 \\ 
L2.0 & 0.02/0.04/0.06  & 0.02/0.05/0.08 \\ 
L2.1 & 0.01/0.02/0.07  & 0.01/0.01/0.02 \\ 
Linf.0 & 0.04/0.10/0.11  & 0.03/0.09/0.14 \\ 
Linf.1 & 0.01/0.02/0.04  & 0.01/0.03/0.06 \\ 
pi.L1.1 & 0.02/0.07/0.16  & 0.05/0.07/0.11 \\ 
pi.L2.1 & 0.02/0.07/0.14  & 0.05/0.06/0.16 \\ 
pi.Linf.1 & 0.01/0.04/0.09  & 0.01/0.06/0.12 \\ 

\\\bottomrule
    \end{tabular}
    \caption{Rejection rates for the FWER procedure as applied to randomly selected collections of acyclic point clouds. We have excluded the cross-polytope acyclic models due to their surprisingly high rejection rates already in the single-hypothesis case. Rates for $\alpha=0.01/0.05/0.10$ are separated with $/$ in each table cell.}
    \label{tab:rejection-fwer-acyclic}
\end{table}

\subsection{Power of the FWER control method}

To estimate statistical power the FWER control method, we do the following experiment 100 times for each of the power estimation model types:

\begin{enumerate}
    \item Draw a number $k$ of observed point clouds at random from a Poisson distribution with mean 10.
    \item Draw one power estimation model at random from the given model type.
    \item Draw $k-1$ acyclic models at random from all the models we have defined.
    \item For each estimator (axis and unbiased hull) draw 100 null model point clouds for each observed point cloud, and compute standardized topological invariants for each.
    \item Maximize invariants, rank and compute p-values.
\end{enumerate}

\begin{figure}
    \centering
    \includegraphics[width=\linewidth]{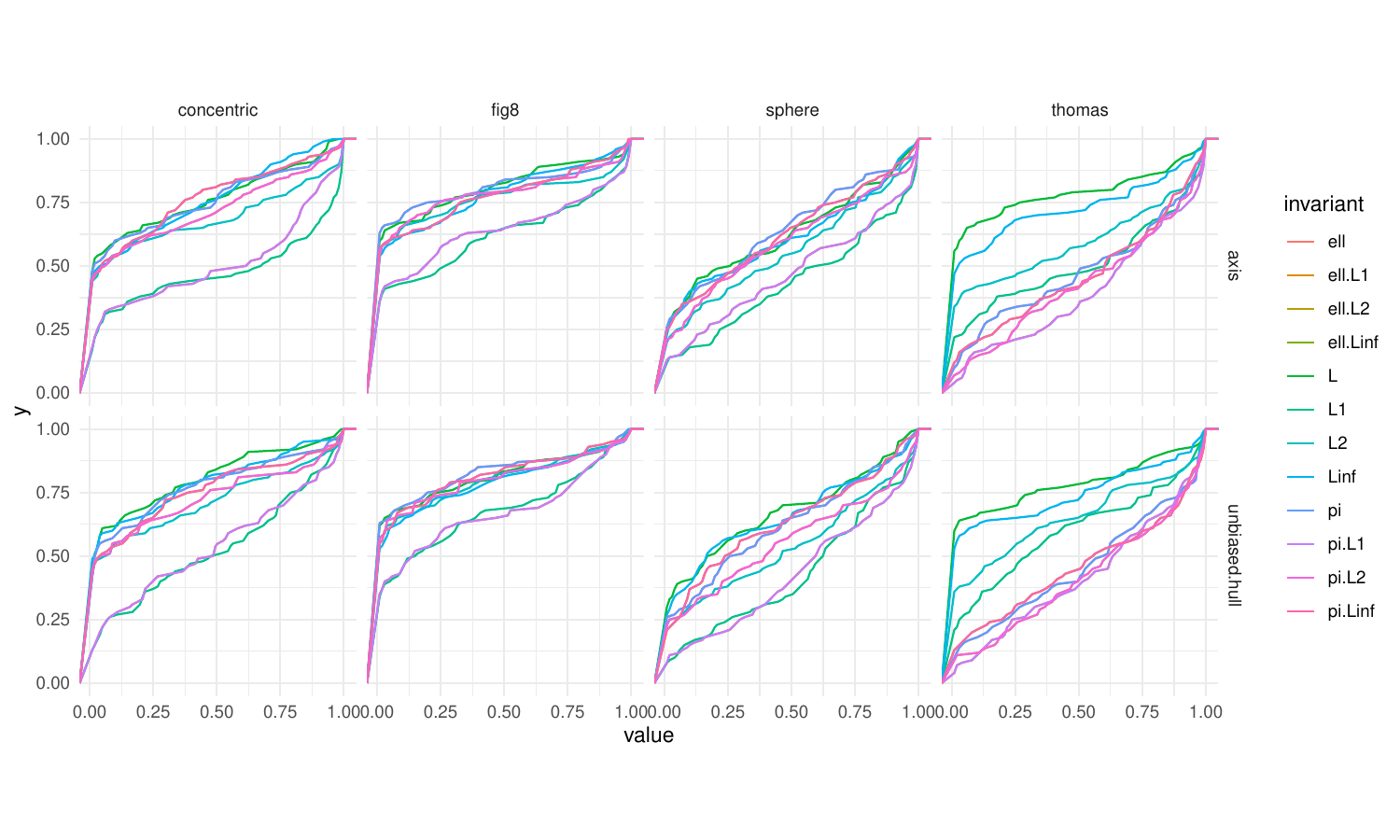}
    \caption{ECDF-plots for the FWER procedure on collections of acyclic point clouds with a single power estimation point cloud, split by type of power estimation.}
    \label{fig:ecdf-fwer-power}
\end{figure}

The resulting p-values are displayed in ECDF-plots in Figure \ref{fig:ecdf-fwer-power}, and corresponding rejection rates for $\alpha=0.01/0.05/0.10$ are reported in Table \ref{tab:rejection-fwer-power}.

\begin{table}
    \centering
    \begin{tabular}{llcc}
    \toprule
        Model type & Invariant & axis & unbiased hull  \\ \midrule
\multirow{12}{*}{concentric} & ell.L1.1 & 0.00/0.00/0.00  & 0.00/0.00/0.00 \\ 
 & ell.L2.1 & 0.05/0.15/0.21  & 0.09/0.14/0.22 \\ 
 & ell.Linf.1 & 0.36/0.43/0.50  & 0.31/0.41/0.47 \\ 
 & L1.0 & 0.02/0.04/0.06  & 0.03/0.05/0.08 \\ 
 & L1.1 & 0.23/0.24/0.29  & 0.17/0.29/0.31 \\ 
 & L2.0 & 0.02/0.09/0.10  & 0.02/0.04/0.07 \\ 
 & L2.1 & 0.44/0.49/0.51  & 0.44/0.48/0.52 \\ 
 & Linf.0 & 0.29/0.33/0.38  & 0.29/0.34/0.35 \\ 
 & Linf.1 & 0.47/0.52/0.57  & 0.44/0.47/0.54 \\ 
 & pi.L1.1 & 0.18/0.25/0.30  & 0.18/0.31/0.37 \\ 
 & pi.L2.1 & 0.30/0.34/0.41  & 0.27/0.37/0.42 \\ 
 & pi.Linf.1 & 0.42/0.48/0.50  & 0.40/0.46/0.51 \\ 
\midrule
\multirow{12}{*}{fig8} & ell.L1.1 & 0.00/0.00/0.00  & 0.00/0.00/0.00 \\ 
 & ell.L2.1 & 0.03/0.10/0.14  & 0.04/0.10/0.15 \\ 
 & ell.Linf.1 & 0.32/0.34/0.37  & 0.28/0.36/0.38 \\ 
 & L1.0 & 0.03/0.05/0.06  & 0.01/0.01/0.01 \\ 
 & L1.1 & 0.42/0.45/0.49  & 0.38/0.42/0.46 \\ 
 & L2.0 & 0.03/0.06/0.07  & 0.01/0.03/0.04 \\ 
 & L2.1 & 0.54/0.62/0.66  & 0.55/0.60/0.64 \\ 
 & Linf.0 & 0.04/0.16/0.23  & 0.14/0.22/0.27 \\ 
 & Linf.1 & 0.59/0.64/0.66  & 0.60/0.65/0.65 \\ 
 & pi.L1.1 & 0.36/0.44/0.53  & 0.38/0.47/0.56 \\ 
 & pi.L2.1 & 0.39/0.51/0.59  & 0.46/0.54/0.63 \\ 
 & pi.Linf.1 & 0.53/0.58/0.60  & 0.54/0.58/0.64 \\ 
\midrule
\multirow{12}{*}{sphere} & ell.L1.1 & 0.00/0.00/0.00  & 0.00/0.00/0.00 \\ 
 & ell.L2.1 & 0.06/0.14/0.18  & 0.04/0.16/0.27 \\ 
 & ell.Linf.1 & 0.19/0.29/0.35  & 0.22/0.31/0.34 \\ 
 & L1.0 & 0.00/0.01/0.02  & 0.06/0.09/0.11 \\ 
 & L1.1 & 0.16/0.19/0.24  & 0.13/0.15/0.20 \\ 
 & L2.0 & 0.01/0.04/0.05  & 0.03/0.05/0.12 \\ 
 & L2.1 & 0.27/0.32/0.36  & 0.26/0.29/0.30 \\ 
 & Linf.0 & 0.08/0.16/0.19  & 0.15/0.18/0.20 \\ 
 & Linf.1 & 0.28/0.32/0.35  & 0.27/0.34/0.36 \\ 
 & pi.L1.1 & 0.16/0.26/0.33  & 0.18/0.28/0.35 \\ 
 & pi.L2.1 & 0.28/0.34/0.40  & 0.25/0.36/0.42 \\ 
 & pi.Linf.1 & 0.30/0.36/0.42  & 0.29/0.40/0.46 \\ 
\midrule
\multirow{12}{*}{thomas} & ell.L1.1 & 0.00/0.00/0.00  & 0.00/0.00/0.00 \\ 
 & ell.L2.1 & 0.00/0.04/0.10  & 0.04/0.08/0.14 \\ 
 & ell.Linf.1 & 0.01/0.05/0.14  & 0.05/0.14/0.17 \\ 
 & L1.0 & 0.18/0.21/0.21  & 0.25/0.29/0.33 \\ 
 & L1.1 & 0.07/0.13/0.16  & 0.07/0.07/0.10 \\ 
 & L2.0 & 0.34/0.35/0.38  & 0.45/0.47/0.48 \\ 
 & L2.1 & 0.07/0.11/0.14  & 0.08/0.09/0.11 \\ 
 & Linf.0 & 0.40/0.47/0.51  & 0.47/0.51/0.57 \\ 
 & Linf.1 & 0.07/0.09/0.12  & 0.05/0.10/0.13 \\ 
 & pi.L1.1 & 0.04/0.08/0.17  & 0.03/0.11/0.17 \\ 
 & pi.L2.1 & 0.04/0.10/0.19  & 0.03/0.09/0.18 \\ 
 & pi.Linf.1 & 0.07/0.14/0.21  & 0.07/0.16/0.22 \\ 

\\\bottomrule
    \end{tabular}
    \caption{Rejection rates for the FWER procedure as applied to randomly selected collections of acyclic point clouds, with a single power estimation point cloud included. We have excluded the cross-polytope acyclic models due to their surprisingly high rejection rates already in the single-hypothesis case. Rates for $\alpha=0.01/0.05/0.10$ are separated with $/$ in each table cell.}
    \label{tab:rejection-fwer-power}
\end{table}

We may of course expect the impact of noise on the detection power to be stark. To this end, in Table \ref{tab:rejection-fwer-power-noise} we give overall rejection rates for the \texttt{concentric}, \texttt{fig8} and \texttt{sphere(2)} power models in the plane, split up by noise levels. For all of these models, we expect to see a non-trivial $H_1$.

For these noise level rejection rates, we repeat the procedure above 100 times for each noise level in $\{0.01,0.05,0.1\}$, but restrict the model types to the planar datasets with the corresponding added noise level.

\begin{table}
    \centering
    \begin{tabular}{llcc}
    \toprule
        Noise & Invariant & axis & unbiased hull  \\ \midrule
\multirow{12}{*}{0.01} & ell.L1.1 & 0.00/0.00/0.00  & 0.00/0.00/0.00 \\ 
 & ell.L2.1 & 0.07/0.17/0.19  & 0.06/0.14/0.19 \\ 
 & ell.Linf.1 & 0.43/0.51/0.53  & 0.41/0.44/0.51 \\ 
 & L1.0 & 0.00/0.02/0.04  & 0.03/0.06/0.10 \\ 
 & L1.1 & 0.47/0.54/0.55  & 0.41/0.50/0.54 \\ 
 & L2.0 & 0.01/0.02/0.03  & 0.03/0.04/0.08 \\ 
 & L2.1 & 0.76/0.78/0.79  & 0.73/0.75/0.78 \\ 
 & Linf.0 & 0.27/0.31/0.33  & 0.22/0.28/0.31 \\ 
 & Linf.1 & 0.73/0.78/0.79  & 0.70/0.74/0.74 \\ 
 & pi.L1.1 & 0.43/0.53/0.62  & 0.44/0.55/0.62 \\ 
 & pi.L2.1 & 0.63/0.67/0.74  & 0.66/0.69/0.74 \\ 
 & pi.Linf.1 & 0.59/0.65/0.71  & 0.64/0.68/0.69 \\ 
\midrule
\multirow{12}{*}{0.05} & ell.L1.1 & 0.00/0.00/0.00  & 0.00/0.00/0.00 \\ 
 & ell.L2.1 & 0.05/0.11/0.15  & 0.05/0.11/0.16 \\ 
 & ell.Linf.1 & 0.32/0.42/0.49  & 0.33/0.45/0.48 \\ 
 & L1.0 & 0.00/0.04/0.05  & 0.03/0.09/0.09 \\ 
 & L1.1 & 0.20/0.31/0.34  & 0.17/0.27/0.32 \\ 
 & L2.0 & 0.01/0.05/0.06  & 0.02/0.05/0.08 \\ 
 & L2.1 & 0.47/0.59/0.63  & 0.53/0.60/0.65 \\ 
 & Linf.0 & 0.21/0.26/0.30  & 0.26/0.30/0.34 \\ 
 & Linf.1 & 0.50/0.60/0.64  & 0.52/0.57/0.62 \\ 
 & pi.L1.1 & 0.14/0.26/0.36  & 0.15/0.30/0.35 \\ 
 & pi.L2.1 & 0.26/0.44/0.50  & 0.26/0.43/0.48 \\ 
 & pi.Linf.1 & 0.43/0.49/0.61  & 0.40/0.50/0.60 \\ 
\midrule
\multirow{12}{*}{0.1} & ell.L1.1 & 0.00/0.00/0.00  & 0.00/0.00/0.00 \\ 
 & ell.L2.1 & 0.06/0.08/0.18  & 0.06/0.11/0.20 \\ 
 & ell.Linf.1 & 0.24/0.32/0.35  & 0.23/0.31/0.40 \\ 
 & L1.0 & 0.01/0.04/0.04  & 0.05/0.10/0.11 \\ 
 & L1.1 & 0.18/0.20/0.28  & 0.16/0.22/0.27 \\ 
 & L2.0 & 0.02/0.05/0.05  & 0.03/0.05/0.08 \\ 
 & L2.1 & 0.32/0.35/0.37  & 0.27/0.37/0.41 \\ 
 & Linf.0 & 0.23/0.28/0.32  & 0.22/0.30/0.39 \\ 
 & Linf.1 & 0.35/0.40/0.40  & 0.35/0.43/0.47 \\ 
 & pi.L1.1 & 0.14/0.21/0.25  & 0.15/0.22/0.25 \\ 
 & pi.L2.1 & 0.21/0.27/0.30  & 0.18/0.27/0.34 \\ 
 & pi.Linf.1 & 0.35/0.45/0.49  & 0.36/0.45/0.51 \\ 

\\\bottomrule
    \end{tabular}
    \caption{Rejection rates for the FWER procedure as applied to randomly selected collections of acyclic point clouds, with a single planar power estimation point cloud of a specific noise level included. We have excluded the cross-polytope acyclic models due to their surprisingly high rejection rates already in the single-hypothesis case. Rates for $\alpha=0.01/0.05/0.10$ are separated with $/$ in each table cell.}
    \label{tab:rejection-fwer-power-noise}
\end{table}

\subsection{Using multiple different invariants}

Our final experiments are to illustrate the observations in Section \ref{sec:combining-invariants} by showing validation and statistical power for composite invariants. For each of the $L_1, L_2, L_\infty$ and each three versions of the $\ell$ and $\pi$ invariants, we create a compositve invariant by combining the invariants as evaluated in each homological dimension. We also try combining all the $L_p$ invariants in all dimensions into one composite invariant, all the $\pi$ invariants in all dimensions into one, and all the $\ell$ invariants in all dimensions into a final compositve invariant.

\begin{figure}
    \centering
    \includegraphics[width=\linewidth]{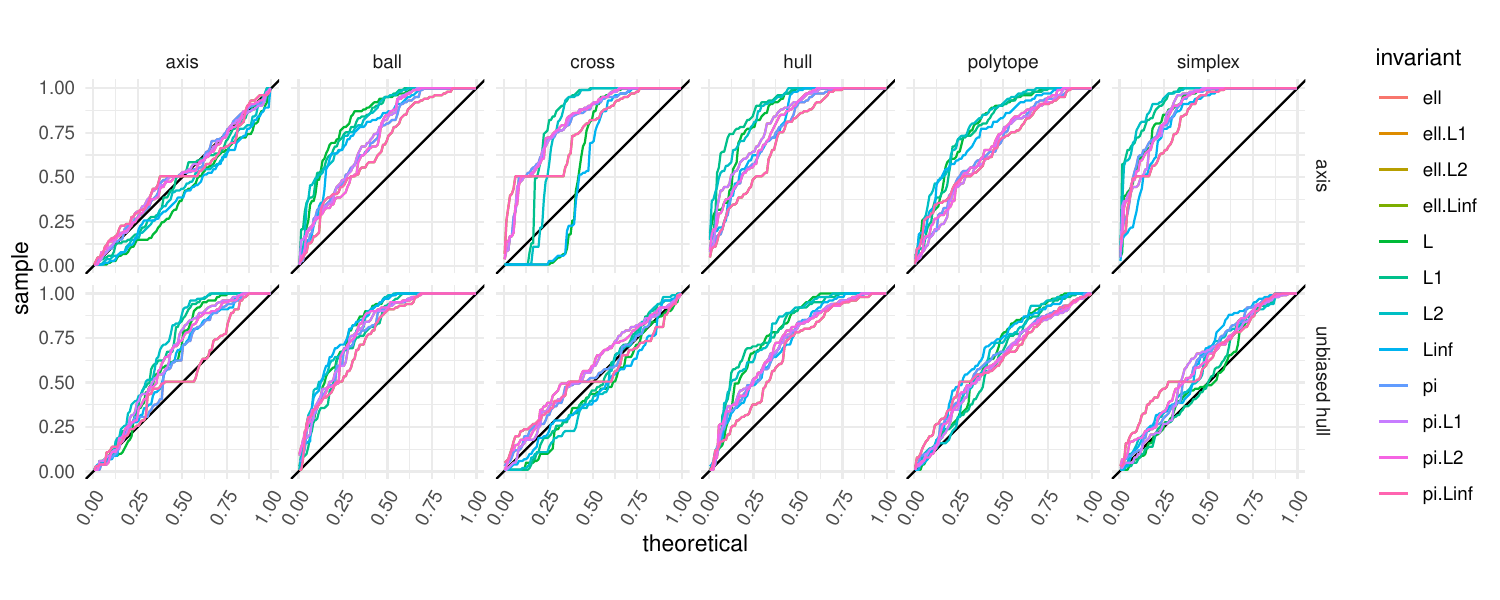}
    \caption{QQ-plots for the single-hypothesis case of all the composite invariants we study here, split up by acyclic model type and null model estimator.}
    \label{fig:qq-composite-validation}
\end{figure}

\begin{figure}
    \centering
    \includegraphics[width=\linewidth]{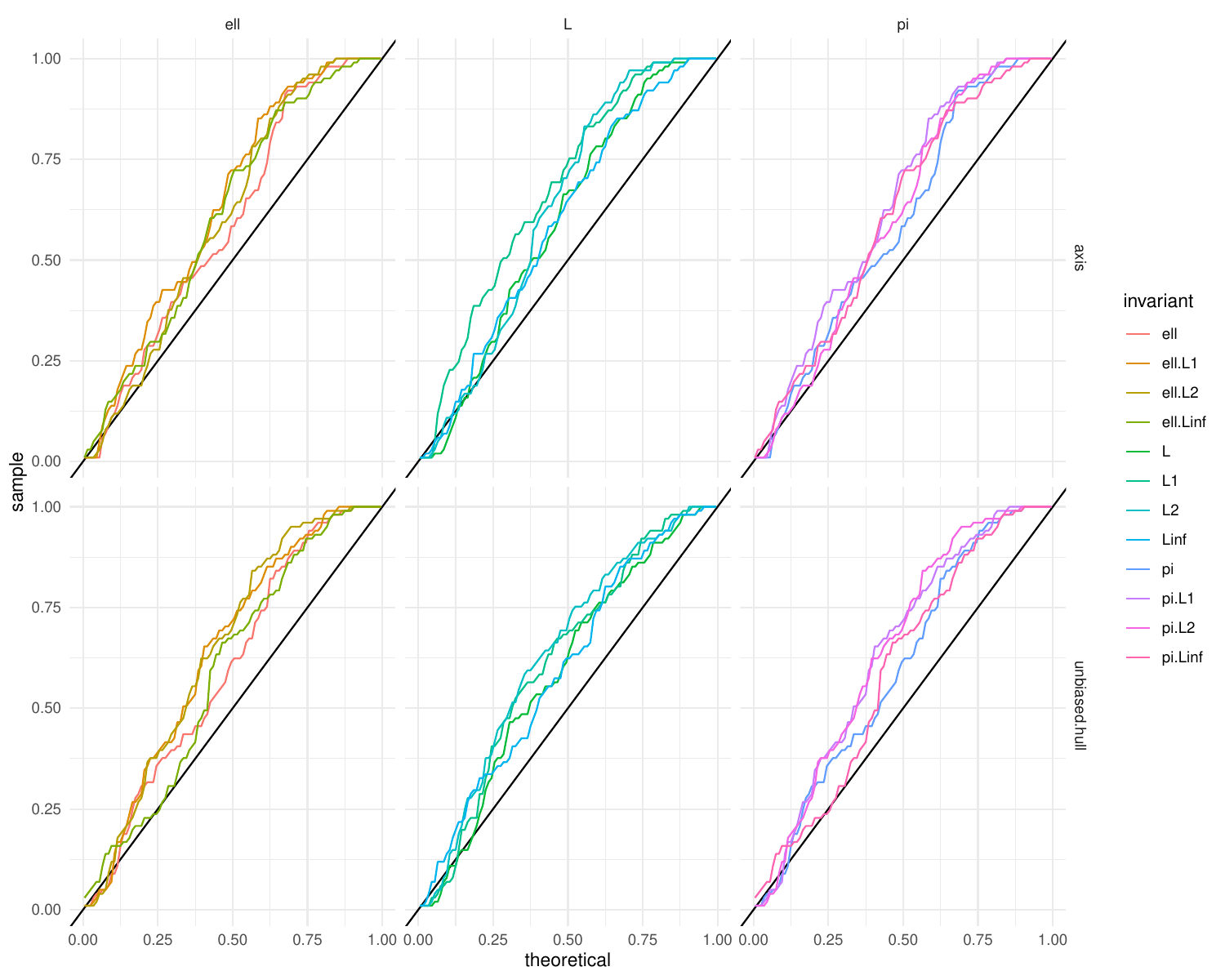}
    \caption{QQ-plots for FWER method for all composite invariants we study here, split up by type ($L_1/L_2/L_\infty$ or $\pi$ or $\ell$) and by null model estimator.}
    \label{fig:qq-fwer-composite-validation}
\end{figure}

In Figure \ref{fig:qq-composite-validation} we show the QQ-plots against a uniform distribution for the single-hypothesis case with composite invariants, and in Figure \ref{fig:qq-fwer-composite-validation}, we show QQ-plots for a multiple hypothesis case.

In Table \ref{tab:reject-rates-acyclic-composite-validation} we show rejection rates for the single-hypothesis composite invariant case, and in Table \ref{tab:reject-rates-acyclic-fwer-composite-validation} we show rejection rates for the multiple-hypothesis composite invariant case. For power estimation, we give rejection rates in Table \ref{tab:reject-rates-power-fwer-composite}.

\begin{longtable}{llcc}
    \toprule
         Model & Invariant & axis & unbiased hull  \\
         \midrule
    \endhead
\multirow{12}{*}{axis}
& \texttt{ell} & 0.02/0.05/0.11 & 0.04/0.08/0.13 \\ 
& \texttt{ell.L1} & 0.02/0.05/0.09 & 0.03/0.03/0.10 \\ 
& \texttt{ell.L2} & 0.03/0.04/0.08 & 0.03/0.09/0.10 \\ 
& \texttt{ell.Linf} & 0.01/0.04/0.06 & 0.01/0.07/0.07 \\ 
& \texttt{L} & 0.08/0.14/0.21 & 0.03/0.04/0.16 \\ 
& \texttt{L1} & 0.02/0.08/0.12 & 0.03/0.03/0.10 \\ 
& \texttt{L2} & 0.05/0.11/0.18 & 0.01/0.05/0.12 \\ 
& \texttt{Linf} & 0.06/0.13/0.19 & 0.01/0.06/0.14 \\ 
& \texttt{pi} & 0.02/0.05/0.11 & 0.04/0.08/0.13 \\ 
& \texttt{pi.L1} & 0.02/0.05/0.09 & 0.03/0.03/0.10 \\ 
& \texttt{pi.L2} & 0.03/0.04/0.08 & 0.03/0.09/0.10 \\ 
& \texttt{pi.Linf} & 0.01/0.04/0.06 & 0.01/0.07/0.07 \\ 
\midrule\multirow{12}{*}{ball}
& \texttt{ell} & 0.01/0.03/0.04 & 0.00/0.00/0.03 \\ 
& \texttt{ell.L1} & 0.01/0.01/0.01 & 0.00/0.00/0.01 \\ 
& \texttt{ell.L2} & 0.01/0.01/0.04 & 0.00/0.00/0.01 \\ 
& \texttt{ell.Linf} & 0.00/0.03/0.05 & 0.02/0.02/0.03 \\ 
& \texttt{L} & 0.01/0.03/0.03 & 0.00/0.01/0.03 \\ 
& \texttt{L1} & 0.01/0.01/0.01 & 0.01/0.01/0.05 \\ 
& \texttt{L2} & 0.01/0.01/0.03 & 0.01/0.01/0.02 \\ 
& \texttt{Linf} & 0.00/0.02/0.03 & 0.00/0.02/0.03 \\ 
& \texttt{pi} & 0.01/0.03/0.04 & 0.00/0.00/0.03 \\ 
& \texttt{pi.L1} & 0.01/0.01/0.01 & 0.00/0.00/0.01 \\ 
& \texttt{pi.L2} & 0.01/0.01/0.04 & 0.00/0.00/0.01 \\ 
& \texttt{pi.Linf} & 0.00/0.03/0.05 & 0.02/0.02/0.03 \\ 
\midrule\multirow{12}{*}{cross}
& \texttt{ell} & 0.00/0.00/0.03 & 0.00/0.03/0.05 \\ 
& \texttt{ell.L1} & 0.00/0.01/0.03 & 0.02/0.07/0.08 \\ 
& \texttt{ell.L2} & 0.00/0.01/0.01 & 0.01/0.02/0.04 \\ 
& \texttt{ell.Linf} & 0.00/0.00/0.01 & 0.00/0.01/0.04 \\ 
& \texttt{L} & 0.26/0.32/0.36 & 0.12/0.17/0.25 \\ 
& \texttt{L1} & 0.14/0.14/0.15 & 0.14/0.16/0.18 \\ 
& \texttt{L2} & 0.15/0.16/0.16 & 0.10/0.15/0.19 \\ 
& \texttt{Linf} & 0.25/0.30/0.35 & 0.02/0.08/0.16 \\ 
& \texttt{pi} & 0.00/0.00/0.03 & 0.00/0.03/0.05 \\ 
& \texttt{pi.L1} & 0.00/0.01/0.03 & 0.02/0.07/0.08 \\ 
& \texttt{pi.L2} & 0.00/0.01/0.01 & 0.01/0.02/0.04 \\ 
& \texttt{pi.Linf} & 0.00/0.00/0.01 & 0.00/0.01/0.04 \\ 
\midrule\multirow{12}{*}{random.hull}
& \texttt{ell} & 0.00/0.00/0.01 & 0.02/0.04/0.04 \\ 
& \texttt{ell.L1} & 0.00/0.00/0.02 & 0.02/0.03/0.05 \\ 
& \texttt{ell.L2} & 0.00/0.00/0.01 & 0.03/0.03/0.05 \\ 
& \texttt{ell.Linf} & 0.00/0.01/0.02 & 0.01/0.02/0.03 \\ 
& \texttt{L} & 0.00/0.00/0.00 & 0.01/0.03/0.04 \\ 
& \texttt{L1} & 0.00/0.00/0.00 & 0.02/0.03/0.04 \\ 
& \texttt{L2} & 0.00/0.00/0.00 & 0.02/0.03/0.05 \\ 
& \texttt{Linf} & 0.00/0.00/0.01 & 0.00/0.02/0.03 \\ 
& \texttt{pi} & 0.00/0.00/0.01 & 0.02/0.04/0.04 \\ 
& \texttt{pi.L1} & 0.00/0.00/0.02 & 0.02/0.03/0.05 \\ 
& \texttt{pi.L2} & 0.00/0.00/0.01 & 0.03/0.03/0.05 \\ 
& \texttt{pi.Linf} & 0.00/0.01/0.02 & 0.01/0.02/0.03 \\ 
\midrule\multirow{12}{*}{random.polytope}
& \texttt{ell} & 0.00/0.01/0.03 & 0.00/0.03/0.07 \\ 
& \texttt{ell.L1} & 0.00/0.04/0.06 & 0.01/0.05/0.08 \\ 
& \texttt{ell.L2} & 0.00/0.02/0.05 & 0.01/0.04/0.11 \\ 
& \texttt{ell.Linf} & 0.01/0.01/0.03 & 0.00/0.01/0.04 \\ 
& \texttt{L} & 0.00/0.00/0.02 & 0.04/0.06/0.09 \\ 
& \texttt{L1} & 0.00/0.02/0.05 & 0.01/0.05/0.08 \\ 
& \texttt{L2} & 0.00/0.02/0.03 & 0.00/0.04/0.06 \\ 
& \texttt{Linf} & 0.01/0.02/0.02 & 0.03/0.06/0.08 \\ 
& \texttt{pi} & 0.00/0.01/0.03 & 0.00/0.03/0.07 \\ 
& \texttt{pi.L1} & 0.00/0.04/0.06 & 0.01/0.05/0.08 \\ 
& \texttt{pi.L2} & 0.00/0.02/0.05 & 0.01/0.04/0.11 \\ 
& \texttt{pi.Linf} & 0.01/0.01/0.03 & 0.00/0.01/0.04 \\ 
\midrule\multirow{12}{*}{simplex}
& \texttt{ell} & 0.00/0.00/0.01 & 0.00/0.02/0.08 \\ 
& \texttt{ell.L1} & 0.00/0.01/0.01 & 0.00/0.08/0.12 \\ 
& \texttt{ell.L2} & 0.00/0.00/0.01 & 0.00/0.01/0.05 \\ 
& \texttt{ell.Linf} & 0.00/0.00/0.00 & 0.00/0.03/0.03 \\ 
& \texttt{L} & 0.00/0.01/0.02 & 0.05/0.08/0.15 \\ 
& \texttt{L1} & 0.00/0.01/0.01 & 0.01/0.10/0.14 \\ 
& \texttt{L2} & 0.00/0.00/0.00 & 0.03/0.05/0.07 \\ 
& \texttt{Linf} & 0.00/0.01/0.01 & 0.02/0.05/0.07 \\ 
& \texttt{pi} & 0.00/0.00/0.01 & 0.00/0.02/0.08 \\ 
& \texttt{pi.L1} & 0.00/0.01/0.01 & 0.00/0.08/0.12 \\ 
& \texttt{pi.L2} & 0.00/0.00/0.01 & 0.00/0.01/0.05 \\ 
& \texttt{pi.Linf} & 0.00/0.00/0.00 & 0.00/0.03/0.03 \\ 
\midrule
\\\bottomrule
    \caption{Single hypothesis rejection rates by acyclic model, estimator and composite invariant for each type of acyclic model. Rates for $\alpha=0.01/0.05/0.10$ are separated with $/$ in each table cell.}
    \label{tab:reject-rates-acyclic-composite-validation}
\end{longtable}

\begin{table}
    \centering
    \begin{tabular}{lcc}
    \toprule
         Invariant & axis & unbiased hull  \\
         \midrule
ell & 0.06/0.06/0.09  & 0.03/0.08/0.11 \\ 
ell.L1 & 0.05/0.06/0.07  & 0.03/0.08/0.10 \\ 
ell.L2 & 0.04/0.07/0.09  & 0.04/0.08/0.09 \\ 
ell.Linf & 0.01/0.04/0.07  & 0.00/0.03/0.06 \\ 
L & 0.05/0.10/0.12  & 0.05/0.08/0.10 \\ 
L1 & 0.05/0.06/0.06  & 0.04/0.07/0.12 \\ 
L2 & 0.04/0.07/0.09  & 0.04/0.08/0.10 \\ 
Linf & 0.02/0.06/0.11  & 0.02/0.04/0.06 \\ 
pi & 0.06/0.06/0.09  & 0.03/0.08/0.11 \\ 
pi.L1 & 0.05/0.06/0.07  & 0.03/0.08/0.10 \\ 
pi.L2 & 0.04/0.07/0.09  & 0.04/0.08/0.09 \\ 
pi.Linf & 0.01/0.04/0.07  & 0.00/0.03/0.06 \\ 

\\\bottomrule
    \end{tabular}
    \caption{FWER corrected rejection rates by acyclic model, estimator and composite invariant for each type of acyclic model. Rates for $\alpha=0.01/0.05/0.10$ are separated with $/$ in each table cell.}
    \label{tab:reject-rates-acyclic-fwer-composite-validation}
\end{table}

\begin{longtable}{llcc}
    \toprule
         Model & Invariant & axis & unbiased hull  \\
         \midrule
    \endhead
\multirow{12}{*}{concentric} & ell & 0.48/0.53/0.60  & 0.42/0.55/0.57 \\ 
 & ell.L1 & 0.18/0.29/0.33  & 0.13/0.22/0.26 \\ 
 & ell.L2 & 0.44/0.49/0.54  & 0.43/0.50/0.55 \\ 
 & ell.Linf & 0.45/0.51/0.53  & 0.45/0.51/0.54 \\ 
 & L & 0.48/0.55/0.58  & 0.47/0.61/0.61 \\ 
 & L1 & 0.17/0.28/0.32  & 0.13/0.22/0.26 \\ 
 & L2 & 0.44/0.50/0.54  & 0.42/0.49/0.55 \\ 
 & Linf & 0.46/0.50/0.54  & 0.49/0.59/0.61 \\ 
 & pi & 0.48/0.53/0.60  & 0.42/0.55/0.57 \\ 
 & pi.L1 & 0.18/0.29/0.33  & 0.13/0.22/0.26 \\ 
 & pi.L2 & 0.44/0.49/0.54  & 0.43/0.50/0.55 \\ 
 & pi.Linf & 0.45/0.51/0.53  & 0.45/0.51/0.54 \\ 
\midrule
\multirow{12}{*}{fig8} & ell & 0.63/0.66/0.68  & 0.63/0.65/0.69 \\ 
 & ell.L1 & 0.36/0.42/0.44  & 0.35/0.41/0.45 \\ 
 & ell.L2 & 0.54/0.60/0.65  & 0.57/0.64/0.66 \\ 
 & ell.Linf & 0.57/0.59/0.62  & 0.55/0.61/0.68 \\ 
 & L & 0.60/0.65/0.67  & 0.62/0.66/0.68 \\ 
 & L1 & 0.36/0.41/0.43  & 0.34/0.40/0.45 \\ 
 & L2 & 0.53/0.59/0.64  & 0.57/0.62/0.65 \\ 
 & Linf & 0.54/0.58/0.62  & 0.53/0.61/0.63 \\ 
 & pi & 0.63/0.66/0.68  & 0.63/0.65/0.69 \\ 
 & pi.L1 & 0.36/0.42/0.44  & 0.35/0.41/0.45 \\ 
 & pi.L2 & 0.54/0.60/0.65  & 0.57/0.64/0.66 \\ 
 & pi.Linf & 0.57/0.59/0.62  & 0.55/0.61/0.68 \\ 
\midrule
\multirow{12}{*}{sphere} & ell & 0.26/0.32/0.35  & 0.26/0.28/0.31 \\ 
 & ell.L1 & 0.13/0.14/0.19  & 0.08/0.11/0.15 \\ 
 & ell.L2 & 0.21/0.24/0.30  & 0.23/0.25/0.29 \\ 
 & ell.Linf & 0.25/0.32/0.36  & 0.21/0.25/0.37 \\ 
 & L & 0.24/0.32/0.38  & 0.30/0.39/0.40 \\ 
 & L1 & 0.12/0.14/0.18  & 0.08/0.11/0.16 \\ 
 & L2 & 0.20/0.25/0.29  & 0.23/0.25/0.30 \\ 
 & Linf & 0.23/0.31/0.40  & 0.27/0.35/0.39 \\ 
 & pi & 0.26/0.32/0.35  & 0.26/0.28/0.31 \\ 
 & pi.L1 & 0.13/0.14/0.19  & 0.08/0.11/0.15 \\ 
 & pi.L2 & 0.21/0.24/0.30  & 0.23/0.25/0.29 \\ 
 & pi.Linf & 0.25/0.32/0.36  & 0.21/0.25/0.37 \\ 
\midrule
\multirow{12}{*}{thomas} & ell & 0.10/0.17/0.20  & 0.10/0.16/0.18 \\ 
 & ell.L1 & 0.04/0.08/0.16  & 0.04/0.08/0.11 \\ 
 & ell.L2 & 0.07/0.10/0.13  & 0.09/0.11/0.11 \\ 
 & ell.Linf & 0.12/0.16/0.20  & 0.13/0.17/0.21 \\ 
 & L & 0.56/0.63/0.67  & 0.60/0.64/0.66 \\ 
 & L1 & 0.22/0.24/0.30  & 0.21/0.27/0.32 \\ 
 & L2 & 0.34/0.40/0.42  & 0.36/0.38/0.42 \\ 
 & Linf & 0.47/0.53/0.55  & 0.53/0.59/0.62 \\ 
 & pi & 0.10/0.17/0.20  & 0.10/0.16/0.18 \\ 
 & pi.L1 & 0.04/0.08/0.16  & 0.04/0.08/0.11 \\ 
 & pi.L2 & 0.07/0.10/0.13  & 0.09/0.11/0.11 \\ 
 & pi.Linf & 0.12/0.16/0.20  & 0.13/0.17/0.21 \\ 

\\\bottomrule
    \caption{Rejection rates for the FWER procedure with composite invariants as applied to randomly selected collections of acyclic point clouds, with a single power estimation point cloud included. We have excluded the cross-polytope acyclic models due to their surprisingly high rejection rates already in the single-hypothesis case. Rates for $\alpha=0.01/0.05/0.10$ are separated with $/$ in each table cell.}
    \label{tab:reject-rates-power-fwer-composite} \\
\end{longtable}

\section{Conclusion}
\label{sec:conclusion}

We have proposed testing procedures for coherence with a null model to test for acyclicity (trivial homology), both in a single-hypothesis setting and with control for Family-Wise Error Rate and for False Discovery Rate in both multiple one-sample and multiple two-sample hypothesis test settings.
For all of these, we perform both validation tests (measuring rejection rates on acyclic point clouds) and estimate statistical power (measuring rejection rates on point clouds with known topological structure).

The FDR control methods produce the intended rejection rates by construction, if it is possible with the data at hand.

For FWER control methods, we have empirical validation and power measurement in Section \ref{sec:experiments-results}.

In our computations, we can see low rejection rates for acyclic point clouds (except for the cross-polytope point clouds, for reasons that are not entirely clear to us), and higher rejection rates for point clouds with structure -- with the statistical power falling as noise levels rise.

\subsection{Limitations}
The null model estimation techniques all have significant limitations: the axis-aligned estimation may be a bad fit for the shape of the support, as seen in the cross-polytope case, the biased convex hull method provides a null model on slightly erroneous wcale and both convex hull methods fail completely on data that is contained in a proper affine subspace.

\subsection{Recommendations}
Based on our experiments, in order to determine acyclicity for either a single point cloud or for a whole collection of related point clouds, we recommend:

\begin{itemize}
    \item Use a dimension-agnostic invariant -- ie, a composite invariant that uses all the homological dimensions available.
    \item If you care about scale, use the \texttt{Linf} composite invariant. If you don't care about scale -- \emph{and} you are not seeking information about clustering structures, use the maximum value of the $\pi$ invariants (the \texttt{pi.Linf} composite invariant).
    \item If your data is contained in a linear subspace of positive co-dimension, use the \texttt{axis} null model estimator, otherwise use the unbiased convex hull estimator.
\end{itemize}

\subsection{Future work}
Our simulation experiments do raise some questions. It would be interesting to find out why the rejection rates are so dramatically much higher for the cross-polytope acyclic models than for all the other acyclic models. It would be interesting to see the performance of these methods on concrete problems motivated by real data.

\clearpage

\bibliographystyle{plainurl}
\bibliography{abel}

\appendix

\clearpage
\section{Deriving the bounds for a bounding box}
\label{app:bounding-box}

To derive the bounds of a bounding box for a uniform distribution on an axis-aligned box, we may treat each axis independently.
This reduces the problem to estimating the support $(a,b)$ of a uniform distribution $U(a,b)$.
A complete sufficient statistic for this distribution is $T(X)=(X_{(1)}, X_{(n)}) = (\min_{x\in X} x, \max_{x\in X} x)$.
If $X\sim U(a,b)$, then $Y=(X-a)/(b-a)\sim U(0,1)$ and the order statistics of the standard uniform distribution $U(0,1)$ are known to be $Y_{(k)}\sim\mathrm{Beta}(k,n+1-k)$.
Hence $Y_{(1)}\sim\mathrm{Beta}(1,n)$, $Y_{(n)}\sim\mathrm{Beta}(n,1)$ and $\mathbb{E}[Y_{(1)}]=1/(n+1)$, $\mathbb{E}[Y_{(n)}]=n/(n+1)$.
It follows that 
\[
\mathbb{E}[X_{(1)}]=\frac{b-a}{n+1}+a
\qquad
\mathbb{E}[X_{(n)}]=\frac{n(b-a)}{n+1}+a
\]

Solving this pair of equations for $a$ and $b$, we get:
\[
\hat{a} = \frac{n\mathbb{E}[X_{(1)}]-\mathbb{E}[X_{(n)}]}{n-1}
\qquad
\hat{b} = \frac{n\mathbb{E}[X_{(n)}]-\mathbb{E}[X_{(1)}]}{n-1}
\]

\clearpage
\section{Power-analysis plots}
\label{app:power-plots}

\begin{figure}[hb!p]
    \centering
    \includegraphics[width=\linewidth]{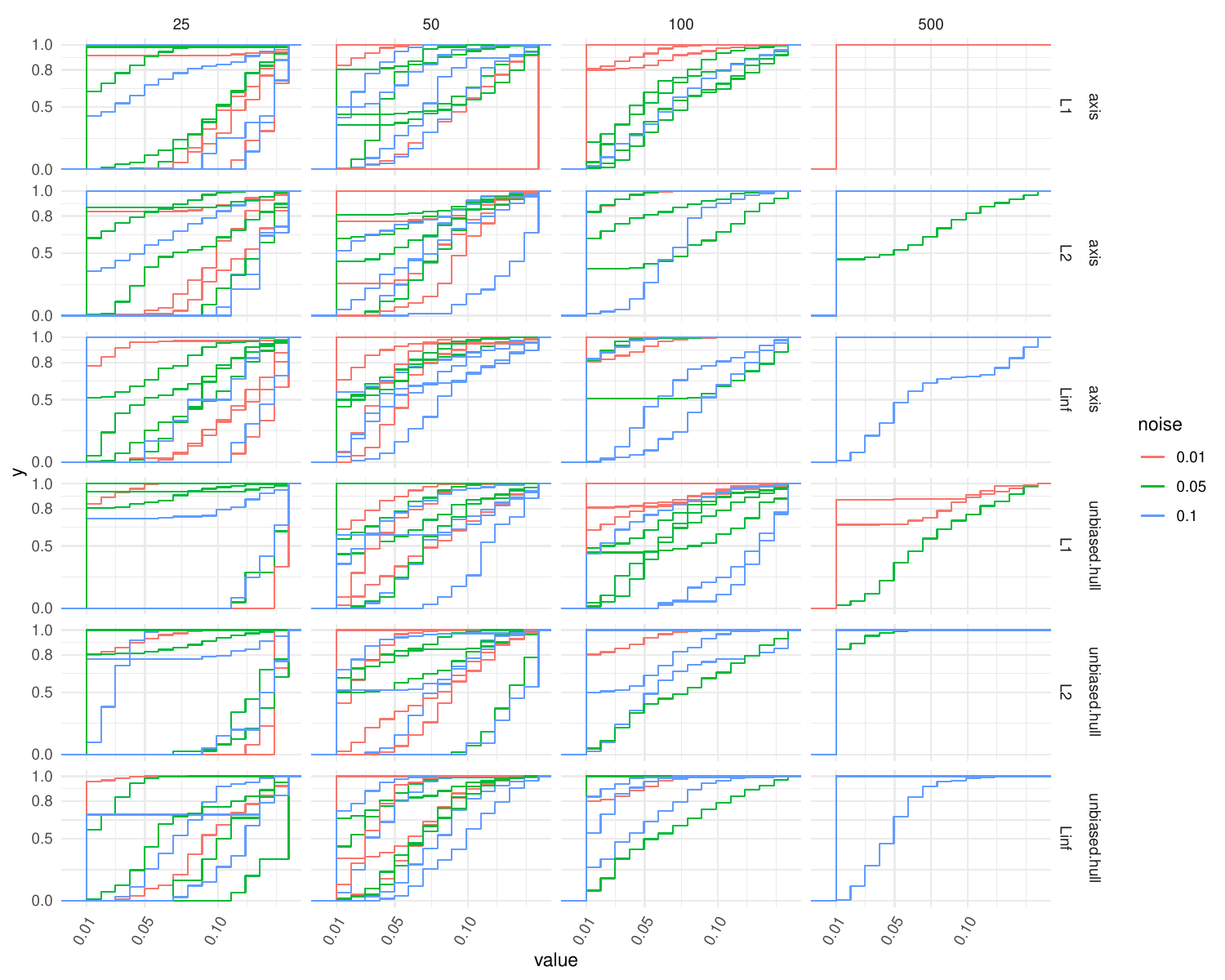}
    \caption{ECDF of the Concentric Circle models across different sample sizes, estimators (axis and unbiased convex hull) and additive topological invariants.}
    \label{fig:power-concentric-ecdf}
\end{figure}

\begin{figure}[hb!p]
    \centering
    \includegraphics[width=\linewidth]{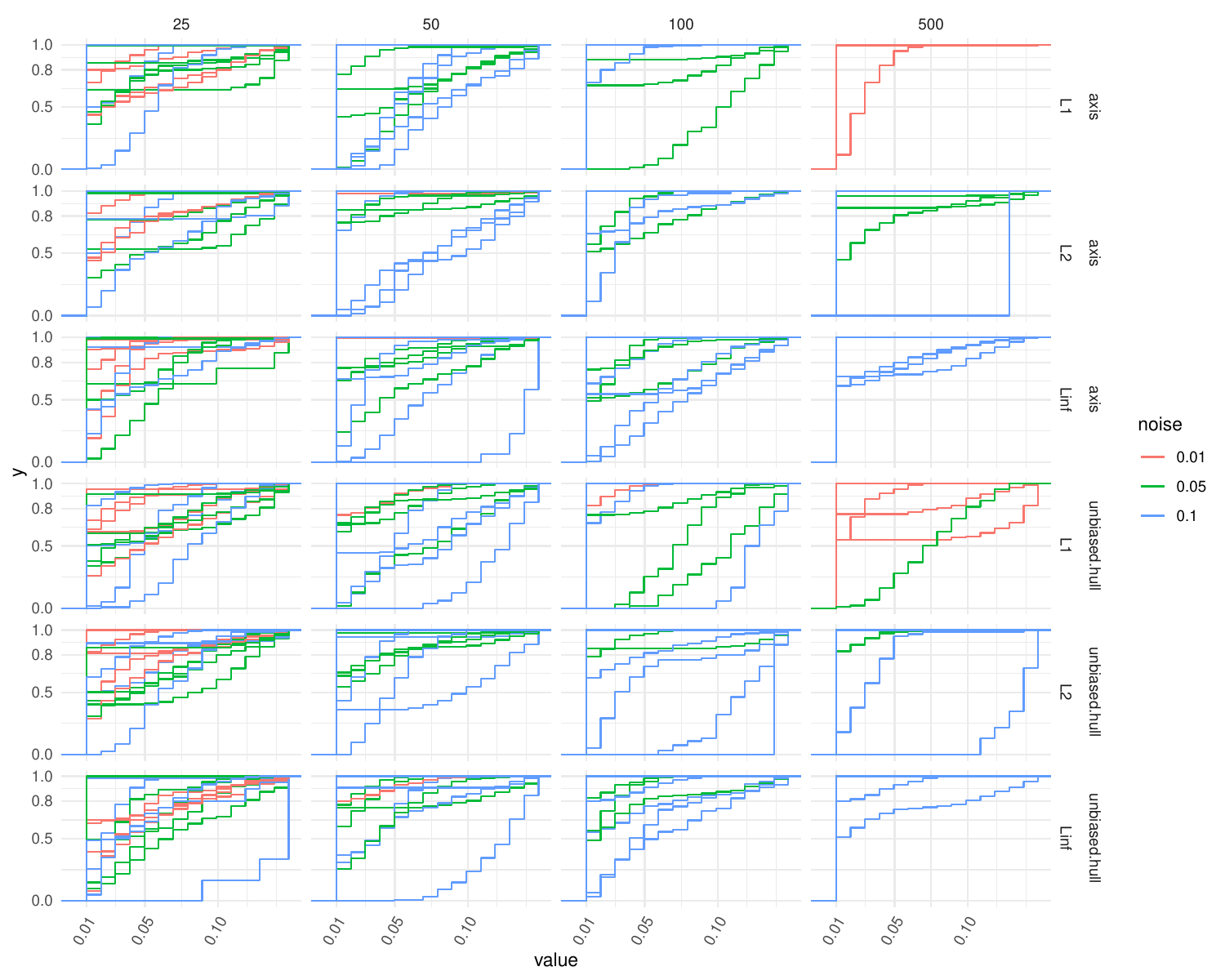}
    \caption{ECDF of the Figure 8 models across different sample sizes, estimators (axis and unbiased convex hull) and additive topological invariants.}
    \label{fig:power-fig8-ecdf}
\end{figure}

\begin{figure}[hb!p]
    \centering
    \includegraphics[width=\linewidth]{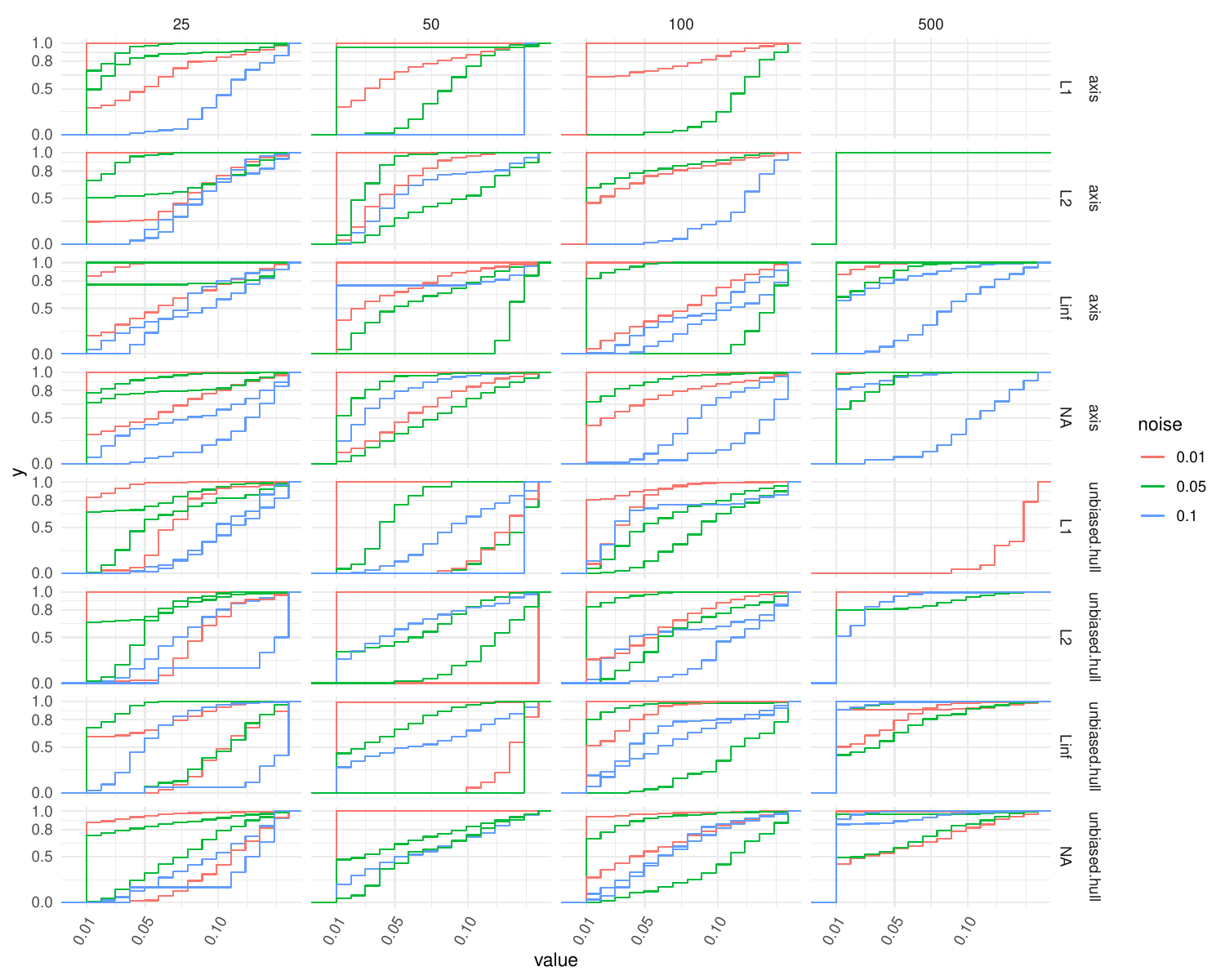}
    \caption{ECDF of the Sphere models across different sample sizes, estimators (axis and unbiased convex hull) and additive topological invariants.}
    \label{fig:power-sphere-ecdf}
\end{figure}

\begin{figure}[hb!p]
    \centering
    \includegraphics[width=\linewidth]{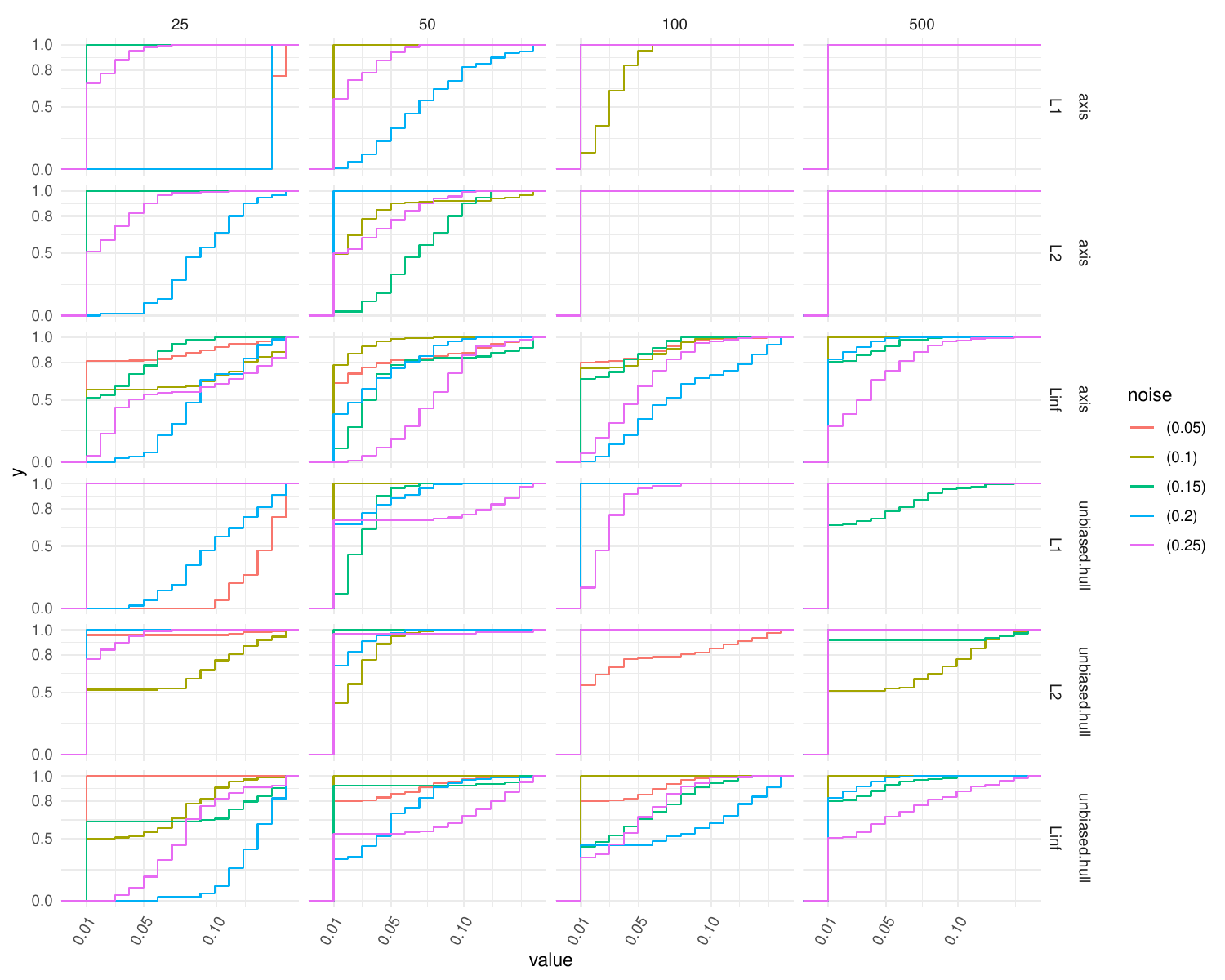}
    \caption{ECDF of the Thomas models across different sample sizes, estimators (axis and unbiased convex hull) and additive topological invariants.}
    \label{fig:power-thomas-ecdf}
\end{figure}

\clearpage
\section{Example point clouds for the alternative models}
\label{sec:point-clouds}

\begin{figure}[hb!p]
    \centering
\foreach\ss in {0.01,0.05,0.1} {
$\sigma^2=\ss$ \\
  \foreach\rr in {1.25,2,5,10} {
  \raisebox{1cm}{$r=\rr$\quad}
    \foreach\NN in {25,50,100,500} {
      \includegraphics[height=1.75cm,trim=0 0 550 0,clip]{pointcloud-figures/power.concentric-\rr-.mvn.\ss-\NN.axis.pdf}
    }
    \\\vspace{-3ex}
  }
}
    \caption{Concentric circles point clouds for a range of noise levels ($\sigma^2$), point cloud sizes (25, 50, 100 and 500 points), and size imbalances ($r$).}
    \label{fig:power-concentric}
\end{figure}

\begin{figure}[hb!p]
    \centering
\foreach\ss in {0.01,0.05,0.1} {
$\sigma^2=\ss$ \\
  \foreach\rr in {0.25,0.5,1,1.5,5} {
  \raisebox{1cm}{$r=\rr$\quad}
    \foreach\NN in {25,50,100,500} {
      \includegraphics[height=1.75cm,trim=0 0 550 0,clip]{pointcloud-figures/power.fig8-\rr-.mvn.\ss-\NN.axis.pdf}
    }
    \\\vspace{-3ex}
  }
}
    \caption{Figure 8 point clouds for a range of noise levels ($\sigma^2$), point cloud sizes (25, 50, 100 and 500 points), and size imbalances ($r$).}
    \label{fig:power-fig8}
\end{figure}

\begin{figure}[hb!p]
    \centering
\foreach\ss in {0.05,0.1,0.15,0.2,0.25} {
$\sigma^2=\ss$ \\
    \foreach\NN in {25,50,100,500} {
      \includegraphics[height=1.75cm,trim=0 0 550 0,clip]{pointcloud-figures/power.thomas-\ss-\NN.axis.pdf}
    }
    \\\vspace{-3ex}
}
    \caption{Thomas process point clouds for a range of noise levels ($\sigma^2$) and point cloud sizes (25, 50, 100 and 500 points).}
    \label{fig:power-thomas}
\end{figure}

\clearpage
\section{Rejection rates for all our models}

For each model specification, we create a bootstrap estimate by repeating 20 times the drawing one of the 5 realizations of that model specification at random, and then draw 100 estimated null model point clouds for each estimation method to compute a p-value for that realizations. The resulting rates are in the supplemental file \texttt{reject-rates-individual-fods.csv}, and the simulated p-values in the supplemental file \texttt{p-values-individual-fods.csv}.

\end{document}